\documentclass[aps,prl,reprint,groupedaddressb,superscriptaddress,twocolumn]{revtex4-2}

\usepackage{graphicx}
\usepackage{color}

\usepackage{bm}
\usepackage{float}
\usepackage{amsthm,amssymb,amsmath,url,braket,amsbsy} 
\usepackage{threeparttable}
\usepackage[utf8]{inputenc}
\usepackage{dirtytalk}
\graphicspath {{./Figure/}}
\usepackage{color}
\usepackage{cancel}

\usepackage{hyperref}
\hypersetup{
    colorlinks=true,
    linkcolor=blue,
    citecolor=blue,
    urlcolor=blue,
}

\usepackage{units}
\usepackage{comment}

\begin{document}
\title{Density matrix renormalization group study of domain wall qubits}

\author{Guanxiong Qu}
\affiliation{RIKEN, Center for Emergent Matter Science (CEMS), Wako-shi, Saitama 351-0198, Japan} 
\author{Ji Zou}
\affiliation{Department of Physics, University of Basel, Klingelbergstr.~82, 4056 Basel, Switzerland}
\author{Daniel Loss}
\affiliation{RIKEN, Center for Emergent Matter Science (CEMS), Wako-shi, Saitama 351-0198, Japan} 
\affiliation{Department of Physics, University of Basel, Klingelbergstr.~82, 4056 Basel, Switzerland}
\affiliation{RIKEN, Center for Quantum Computing (RQC), Wako-shi, Saitama 351-0198, Japan}
\author{Tomoki Hirosawa}
\affiliation{Department of Physical Science, Aoyama Gakuin University, Kanagawa 252-5258, Japan}

\date{\today}

\begin{abstract}
Nanoscale topological spin textures in magnetic systems are emerging as promising candidates for scalable quantum architectures. Despite their potential as qubits, previous studies have been limited to semiclassical approaches, leaving a critical gap: the lack of a fully quantum demonstration. Here, we address this challenge by employing the density-matrix renormalization group (DMRG) method to establish domain wall (DW) qubits in coupled quantum spin-1/2  chains. We calculate the ground-state energies and excitation gaps of the system and find that DWs with opposite chiralities form a well-defined low-energy sector, distinctly isolated from higher excited states in the presence of anisotropies. This renders the chirality states suitable for encoding quantum information, serving as robust qubits. Interestingly, when a magnetic field is applied, we observe  tunneling between  quantum DW states with opposite chiralities. Through quantum simulations, we construct an effective qubit Hamiltonian that exhibits strongly anisotropic $g$-factors, offering a way to implement single-qubit gates. Furthermore, we obtain an effective interacting Hamiltonian for two mobile DWs in coupled quantum spin chains from DMRG simulations, enabling the implementation of two-qubit gates.Single-qubit and two-qubit gates are also demonstrated in real-time simulations using the time-dependent variational principle.
Our work represents a critical step from semiclassical constructions to a fully quantum demonstration of the potential of DW textures for scalable quantum computing, establishing a solid foundation for future quantum architectures based on topological magnetic textures.
\end{abstract}
\maketitle
\date{\today}

\section{I. Introduction}
Magnetic domain walls~(DWs) are quasi-one-dimensional topological defects characterized by a winding number~\cite{zang2018topology}. Their topologically protected spin textures provide robustness, making DWs highly attractive for classical information applications~\cite{nataf2020domain}. This potential has spurred significant experimental efforts, leading to recent advancements in the efficient manipulation of DWs with sub-$\unit[10]{nm}$ widths at velocities exceeding $\unit[100]{m/s}$~\cite{yang2015domain,kim_fast_2017}.  Leveraging mobile DWs as information carriers, racetrack memory offers a promising route to fast and high-density data storage~\cite{parkin2008magnetic}.

Beyond classical applications, nanoscale spin textures provide an exciting platform for macroscopic quantum phenomena at low temperatures~\cite{braun_berrys_1996}. With increasing interest in expanding quantum computing platforms, recent proposals have explored the use of nanoscale magnetic textures, such as skyrmions~\cite{psaroudaki_skyrmion_2021,xia_universal_2023} and DWs~\cite{takei_quantum_2018,zou_quantum_2023}, as qubits. These robust and high-speed topological textures show great potential for functioning as flying qubits~\cite{divincenzo2000physical,yamamoto_electrical_2012,zou_quantum_2023,zou2024topological}. In addition, single- and two-qubit gates have been proposed via external fields and current-driven motion~\cite{zou_quantum_2023}, laying the foundation for universal quantum computation within existing spintronic technologies. However, while the potential of topological textures as qubits hinges fundamentally on their quantum nature, all previous proposals have been limited to semiclassical treatments, valid for large spin, using collective-coordinate approaches within the continuum limit of the model. This situation is highly unsatisfactory, since  spins are intrinsically quantum objects with their quantum properties being crucial to enabling quantum computation in such systems. A fully quantum treatment for spin-1/2, critical for rigorously establishing the feasibility of topological textures as robust qubits, remains an open and pressing challenge.

 \begin{figure}[t]
 \includegraphics[width=8cm]{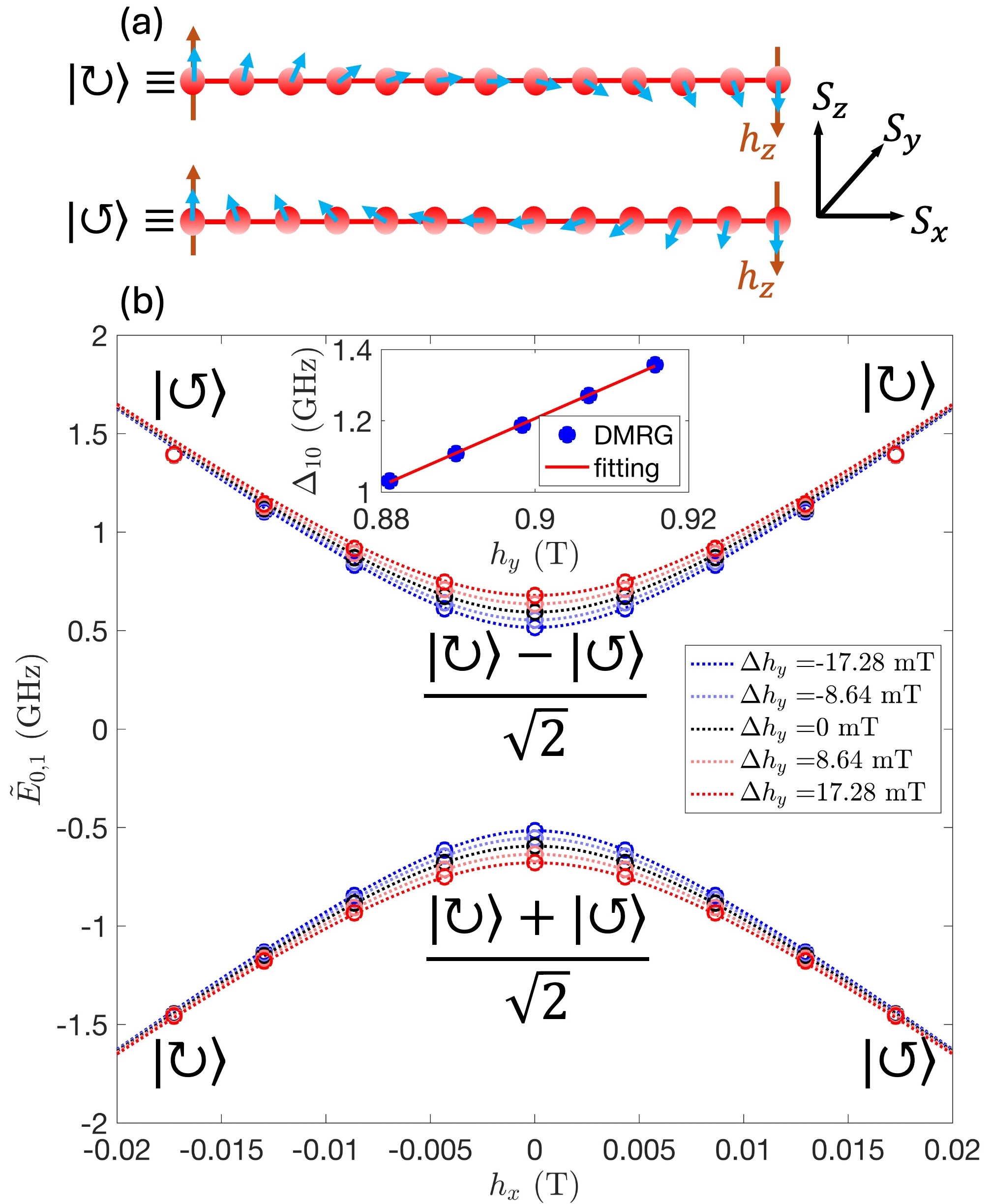}
 \caption{\label{fig:1} (a)~Schematic illustration of DW chirality states in a single quantum DW. (b)~DMRG simulation data (circle) and effective Hamiltonian (dotted line) of DW qubits energy spectra with tuning of external magnetic fields $h_{x,y}$. The inset shows tuning of the qubit splitting $\Delta_{10}$ by small variation of $h_y$. $\tilde{E}_{0,1}$ are energies shifted such that the zero-energy level is aligned with the center of the qubit splitting. Parameters for DMRG simulations are shown in Tab.~\ref{tab:1}. }
 \end{figure}
In this paper, we address this critical challenge by employing the density-matrix renormalization group (DMRG) method~\cite{white_density_1992,schollwock_density-matrix_2005, 10.21468/SciPostPhysCodeb.4} to investigate quantum properties of ferromagnetic DW (kink-type) textures in coupled quantum spin chains~\cite{schilling_quantum_1977}. We focus on the chirality of domain walls, defined as the winding direction of spins across the ferromagnetic domain wall. By examining the ground-state and low-lying excited-state properties of these DW textures, we find that DW states with opposite chiralities, as depicted in Fig.~\ref{fig:1}(a), form a well-defined low-energy sector that is well-separated from higher excited states. This isolation enables the encoding of quantum information into the DW chiralities. Remarkably, we observe that the tunneling rate and energy detuning between the two  quantum DW chirality states are highly tunable via magnetic fields applied along the $y$- and $x$-directions, respectively, characterized by an anisotropic effective $g$-factor. Furthermore, by restricting to the chirality subspace, we construct an effective Hamiltonian for a single DW qubit, capturing its essential quantum dynamics. Moreover, to demonstrate the two-qubit gate between DWs, we consider two coupled quantum spin chains. By calculating the ground-state and excited-state energies for DWs at varying separations, we construct an effective Hamiltonian for the mobile DWs, enabling the implementation of two-qubit gates. 
Our work establishes the feasibility of utilizing DW textures for universal quantum computation within a fully quantum framework, surpassing previous studies that relied on semiclassical approaches. This provides a rigorous and solid foundation for leveraging topological textures as promising platforms for future quantum architectures.

\section{II. DMRG simulation for quantum DW}
We calculate the ground-state and low-lying excited state energies of a DW texture in an open anisotropic (XYZ) ferromagnetic quantum spin-$1/2$  chain with $N$ sites, governed by the following Hamiltonian:
\begin{align}
H_{\text{chain}} &= \sum^{N-1}_{i=1} \left( -J \bm{S}_i \cdot \bm{S}_{i+1} -K_z  S^z_i  S^z_{i+1} + K_y S^y_i  S^y_{i+1} \right) \notag \\
                &+ \sum^{N}_{i=1} \left(\mu_B h_{x} S^x_i + \mu_B h_{y} S^y_i \right)\! +\! \mu_B h_z(S^z_1\! -\! S^z_N) ,
\label{eq:1}
\end{align}
where $S^{\alpha}_i$ denotes the spin-$1/2$  operators with $\alpha\in\{x,y,z\}$, $J>0$ represents isotropic ferromagnetic coupling, and $K_{z,y}>0$ denote easy $z$-axis and hard $y$-axis exchange anisotropies, respectively. Here, $h_{x,y}$ is an external magnetic field along the $x$- or $y$-axis. To focus on the DW spin texture, the boundary spins ($i = 1, N$) are  coupled to pinning $z$-fields ($h_z \gg J, K_{z,y}$), ensuring a kink-type ground state~\cite{schilling_quantum_1977, alcaraz_anisotropic_1995, alcaraz_entanglement_2004}. Such pinning fields can emerge from exchange coupling the boundary spins to magnets with magnetization along  $\pm z$ direction. We remark that,  in the absence of hard axis anisotropy and external magnetic fields $(K_y=0,h_{x,y}=0)$, Eq.~(\ref{eq:1}) reduces to the XXZ model which can be solved  analytically with fixed boundary condition~\cite{alcaraz_anisotropic_1995} and anti-periodic boundary condition~\cite{BRAUN_DW,PhysRevB.58.5568} to support DWs. However, introducing a finite $K_y$ renders an analytical solution infeasible.

To characterize the chirality of  the DW in a quantum spin chain, we introduce the chirality operators~\cite{tserkovnyak_quantum_2020,PhysRevLett.125.267201,PhysRevResearch.1.033071}  on the spin lattice
\begin{equation}
    C_\gamma = \epsilon_{\alpha \beta \gamma} \sum^{N-1}_{i=1} S_i^{\alpha} S_{i+1}^{\beta},
    \label{eq:2}
\end{equation}
where $\epsilon_{\alpha \beta \gamma}$ is Levi-Civita symbol with $\alpha,\beta,\gamma\in\{x,y,z\}$. By evaluating the energies of the first few states and their chiralities, we find that, for $K_y>0$, the ground state exhibits degenerate DW chiralities and importantly is well-separated from higher-energy states. For instance, with $K_y = 0.1$~meV, the gap exceeds $9.6$~GHz and remains largely unaffected by the  $h_y$ field, even up to 1~T~\cite{supplementary}. This large gap between the chirality-state subspace and higher excited states enables the  encoding of quantum information within this well-isolated subspace (qubit space). We note that in the XXZ limit the ground state is non-degenerate for $N$ even while doubly degenerate for $N$ odd~\cite{schilling_quantum_1977,PhysRevLett.122.097204}. However, already a small easy-axis anisotropy $K_y$ sufficiently  suppresses these finite-size parity effects~\cite{supplementary}.

We then investigate the effects of an in-plane magnetic field on the low-energy states using DMRG. The parameters used for the  simulations are summarized in Tab.~\ref{tab:1}. We find that applying a strong magnetic field along the $y$-direction (e.g., $h_y = 0.9\,\text{T}$) lifts the degeneracy of the chirality states, resulting in a  $\Delta_{10} \equiv E_1 - E_0$ in the GHz regime, as shown in the inset of Fig.~\ref{fig:1}(b). Physically, this magnetic field suppresses the tunneling barrier and facilitates quantum tunneling between the two chirality states~\cite{braun_berrys_1996,zou_quantum_2023}. Additionally, in Fig.~\ref{fig:1}(b), we demonstrate the tunability of the qubit splitting through both a small $h_x$ field and slight variations in the $h_y$ field near the bias field $h_{y,\text{bias}} = 0.9\,\text{T}$.

 \begin{table}[t]
\caption{\label{tab:1}  Parameters for DW qubits on coupled spin chains.  We use $N=30$, $J = 25.85$ meV, $K_z=0.26$ meV, $K_y=0.1$ meV, $\mu_B h_z=-100$ meV, and $h_y=0.9$ T for DMRG simulations. }
\begin{ruledtabular}
\begin{tabular}{llll}
$\Delta_{10}$   & $ g_x$ & $g_y$ & $J^{\text{eff}}_{\text{in}}$  \\
  \hline
1.2 GHz &76.2 GHz/T & 9.4 GHz/T & 278 MHz\\
\end{tabular}
\end{ruledtabular}
\end{table}

 \begin{figure}[t]
 \includegraphics[width=8.7cm]{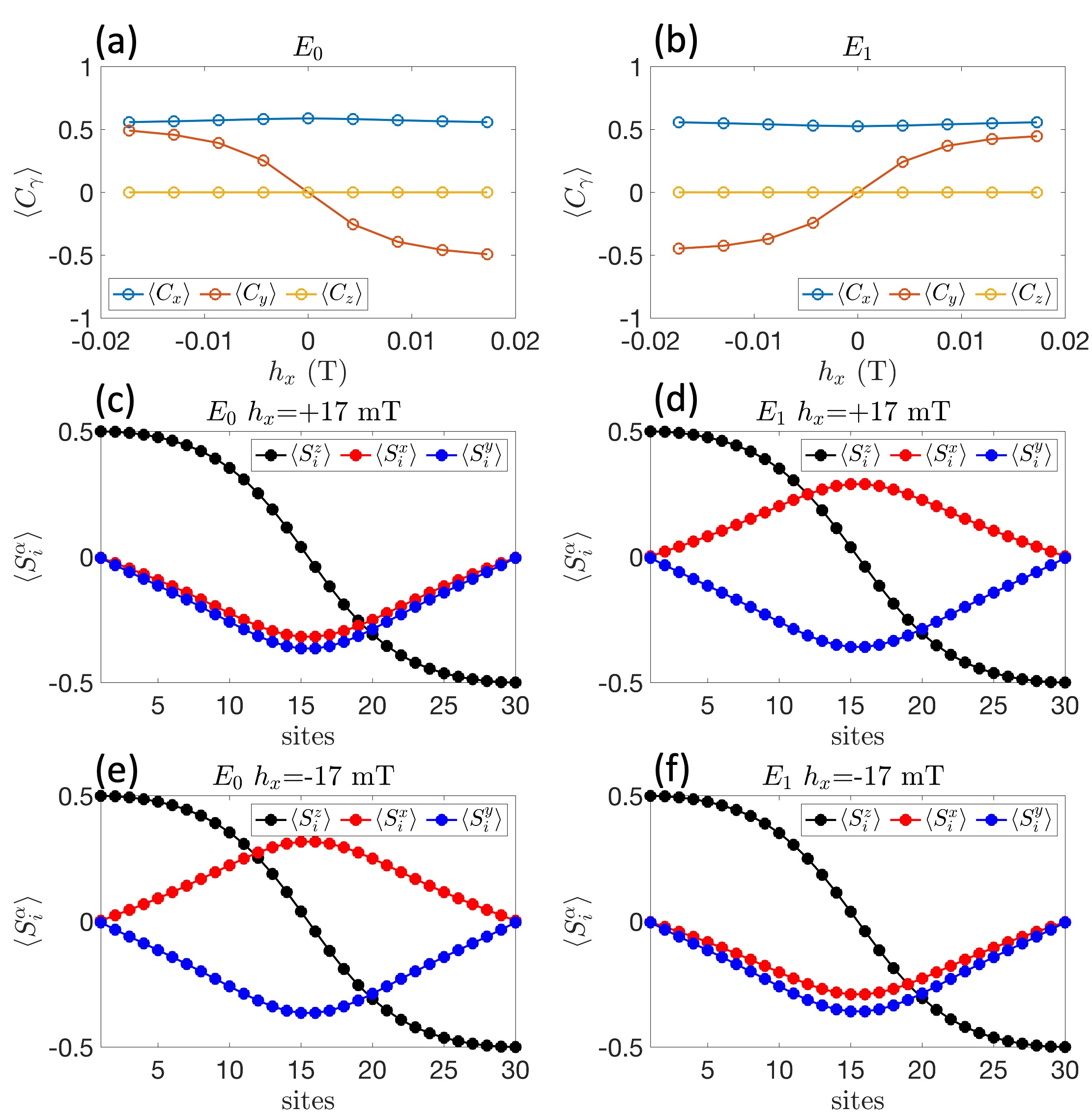}
 \caption{\label{fig:1/2} DW chiralities of a single spin chain with external magnetic fields: (a,b) DW chiralities $\braket{C_\gamma}$ tuning by $h_x$ field; (c,d) spin profiles $S^{x,y,z}_i$ with positive bias field $h_x=+17$ mT; (e,f) spin profiles $S^{x,y,z}_i$ with negative bias field $h_x=-17$ mT. DMRG simulation parameters are the same as in Fig.~\ref{fig:1}. }
 \end{figure}
 
We further show the chirality of the DW qubit in a single quantum chain in Fig.~\ref{fig:1/2}. When tuning the $h_x$ field, we observe that, interestingly, the DW chiralities perpendicular to the hard-axis anisotropy, $C_x$ and $C_z$, remain nearly unchanged. In sharp contrast, the DW chirality $C_y$ shifts with the $h_x$ field. At small bias fields, $h_x = \pm 17$ mT, the qubit states ($E_0$ and $E_1$) exhibit opposite $C_y$-chiralities, denoted as $\ket{\circlearrowright}$ and $\ket{\circlearrowleft}$. Flipping the direction of the magnetic field $h_x$ correspondingly reverses the chirality $C_y$ of the qubit states.
 The spin profiles of a single spin chain, shown in Fig.~\ref{fig:1/2}, reveal the $C_y$-chiralities of DW qubits under small bias fields ($h_x = \pm 17$ mT). Notably, at zero bias field ($h_x = 0$), the qubit states are superposition state of the two $C_y$-chiralities, resulting in a vanishing expectation value of chirality $\braket{C_y}$.

\section{III. Effective Hamiltonian}
We systematically tune the magnetic field $h_{x,y}$ and  find that the qubit splitting $\Delta_{10}$ 
increases exponentially with the $h_y$ field, while it varies linearly with $h_x$~\cite{supplementary}.   From the DMRG simulations,  we construct an effective Hamiltonian for the DW qubit subspace in a single quantum spin chain:
\begin{equation}
H^{\text{eff}}_1 =  \left( \Delta_{10}/2 + g_y \Delta h_y \right)  \sigma_z   - g_x h_x \sigma_x, \
\label{eq:3}
\end{equation}
where $g_{x,y}$ are effective $g$-factors. Here, $ \Delta h_y \equiv h_y - h_{y, \text{bias}}$ denotes a small deviation from the strong bias field, while $\sigma_{x,y,z}$ are pseudo-spin operators defined in the DW chirality subspace spanned by the superposition states $(\ket{\circlearrowright} + \ket{\circlearrowleft})/\sqrt{2}$ and $(\ket{\circlearrowright} - \ket{\circlearrowleft})/\sqrt{2}$. In Fig.~\ref{fig:1}(b), we illustrate the agreement between DMRG simulations and the effective Hamiltonian. The linear dependence of the qubit splitting $\Delta_{10}$ on $h_x$  is clearly observed, allowing the effective $g$-factor to be extrapolated as $g_x \approx 76.2\, \text{GHz/T}$. Similarly, $\Delta_{10}$ varies approximately linearly with $\Delta h_y$, yielding an effective $g$-factor of $g_y \approx 9.4\, \text{GHz/T}$. Thus, a single DW qubit exhibits strongly anisotropic $g$-factors under magnetic field tuning.

These anisotropic $g$-factors in Eq.~(\ref{eq:3}) enable the implementation of Rabi-driving and phase-driving mechanisms for a single DW qubit~\cite{bosco_phase-driving_2023}. Here, we demonstrate Rabi-driven rotation on a single DW qubit by applying an oscillating magnetic field $h_x \cos (\omega_x t)$:
$H^{\text{Rabi}}_1 =  \omega_{q}  \sigma_z/2 - g_x  h_x \cos(\omega_x t) \sigma_x$,
where $\omega_q = \Delta_{10} $ is the qubit frequency. In rotating wave approximation (RWA), the Hamiltonian reads $ \tilde{H}_{1,\text{RWA}} = \Delta \omega  \sigma_z/2 - g_x h_x \sigma_x/2$ with $\Delta \omega = \omega_{q} - \omega_x$. At the resonance condition $\Delta \omega =0$, it yields a single qubit rotation around $x$-axis,  $ U_{1,z}(t) = \exp\left( i \pi g_x h_x \sigma_x t  \right)$,  with Rabi frequency $\Omega_R=2 \pi g_x h_x$. Importantly, we note that the large effective $g$-factor in the $x$ direction enables ultrafast single-qubit gate operations, achieving a gate time of  1~ns with a field amplitude of $h_x \sim 4$~mT. At low field strength $h_x$, the effective Hamiltonian with RWA is in good agreement with real-time simulations of a single DW qubit under linear Rabi driving~\cite{supplementary}.
 
 \begin{figure}[t]
 \centering
 \includegraphics[width = 8.5 cm]{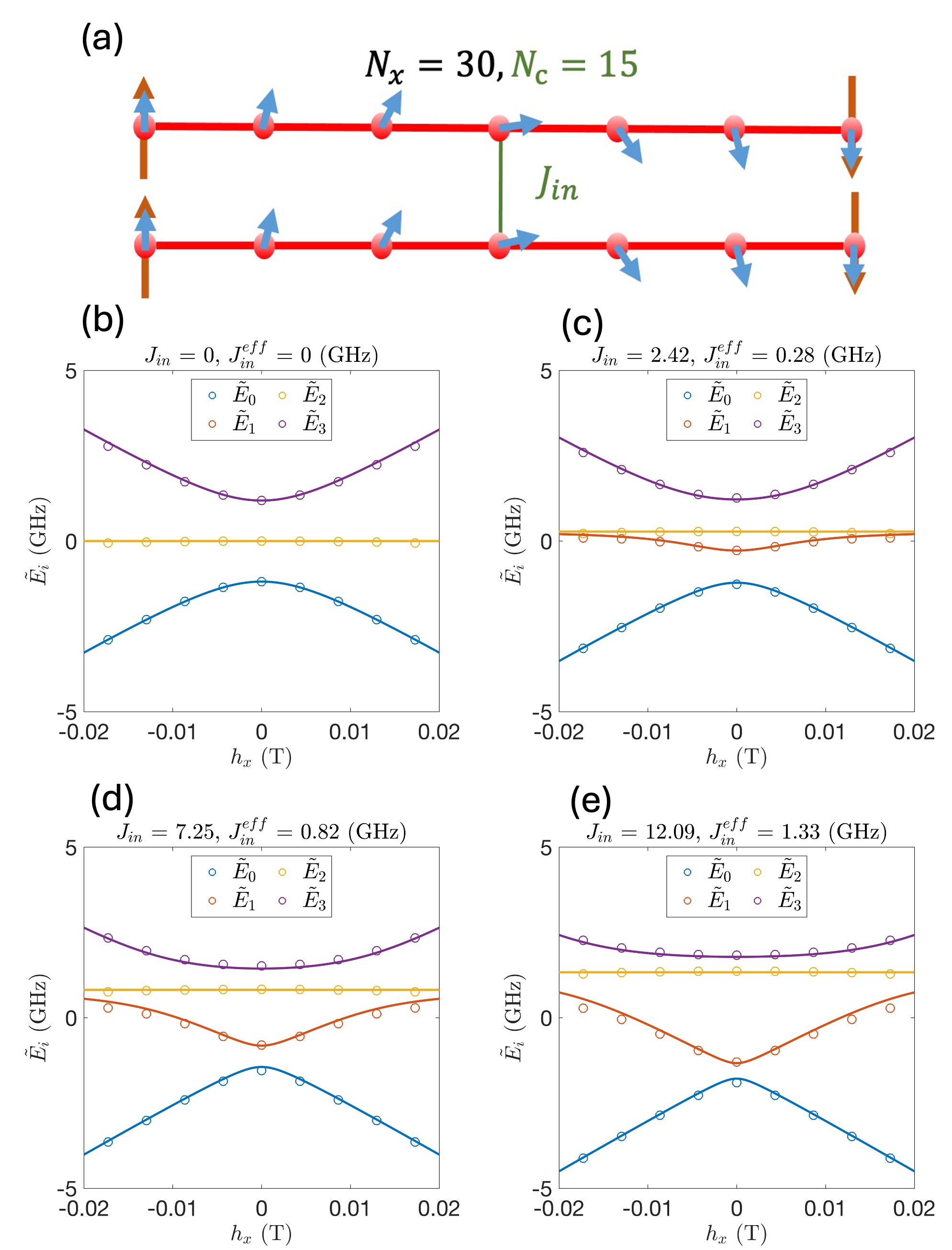}
 \caption{ (a) Schematic illustration of DMRG simulation for  two stationary DW qubits. (b-e) Energy spectra of two-qubit subspace for two stationary DW qubits with various single site coupling strength $J_{\text{in}}= 0,2.42, 7.25, 12.09$ GHz, corresponding to 0, 10, 30, 50 $\mu$eV. Circles represent DMRG data and line plots represent effective Hamiltonian with $J^{\text{eff}}_{\text{in}}= 0, 0.28, 0.82, 1.33$ GHz. Parameters for each DW qubit are the same as in Fig.~\ref{fig:1}.
 }
 \label{fig:3}
 \end{figure}

\section{IV. DMRG simulation for coupled DWs}
We now investigate the interactions between quantum DW textures using DMRG. To this end, we consider a quantum spin ladder~\cite{lauchli_entanglement_2012} composed of two identical spin chains weakly coupled at a single site in the middle of the chains $(N_c = 15)$, as shown in Fig.~\ref{fig:3}(a). Each chain contains a total of $N_x$ sites. We first focus on the coupling between two stationary DW qubits, where the DW centers are pinned to the middle of each chain by $h_z$ [see Fig.\ref{fig:3}(a)]. The spin ladder Hamiltonian is given by:
 \(  H_{\text{ladder}} = H_{\text{chain},1} + H_{\text{chain},2} + H_{\text{in}},  \)
where the interaction Hamiltonian is $H_{\text{in}} = - J_{\text{in}} \bm{S}_{N_c,1} \cdot \bm{S}_{N_c,2}$ and   $H_{\text{chain},1(2)}$ represents the single chain Hamiltonian defined in Eq.~(\ref{eq:1}).  The two spin chains are weakly coupled ferromagnetically at site $N_c$, with $J_{\text{in}}$ satisfying $0 < J_{\text{in}} \ll J$. We point out that the single-site inter-chain coupling is  effectively equivalent to a uniform inter-chain coupling with reduced strength as shown in SM~\cite{supplementary}.

We calculate the energy levels of the first four eigenstates of the two coupled DWs, which form a well-isolated two-qubit subspace. The energy levels as functions of $h_x$ are shown in Figs.~\ref{fig:3}(b-e). We observe that, in the absence of inter-chain coupling [Fig.~\ref{fig:3}(b)], the first excited state ($E_{1,2}$) is degenerate, corresponding to a mixed state of opposite DW chiralities in each spin chain. Importantly, when the inter-chain coupling $J_{\text{in}}$ is introduced, the interaction between the two DW qubits lifts this degeneracy, opening a gap between the $E_1$ and $E_2$ states, as seen in Fig.~\ref{fig:3}(c-e). This interaction between DW qubits enables the implementation of two-qubit gates, which will be discussed in detail later.
From the DMRG simulations,  we extrapolate the effective Hamiltonian for the two coupled DW qubits:
 \begin{equation} 
    H^{\text{eff}}_{2}= H^{\text{eff}}_1 \otimes I_2 +  I_1 \otimes H^{\text{eff}}_2  - J^{\text{eff}}_{\text{in}} \sigma_x \otimes \sigma_x ,
    \label{eq:6}
\end{equation}
where $J^{\text{eff}}_{\text{in}}$ is the effective coupling strength of two DW qubits. The inter-chain coupling induces hybridization of chirality states between the two DW qubits, represented by the term $(\propto \sigma_x \otimes \sigma_x)$ in the effective Hamiltonian. The effective coupling strength $J^{\text{eff}}_{\text{in}}$ scales approximately linearly with $J_{\text{in}}$ in $H_{\text{in}}$. Notably, in the absence of the $h_x$ field, the energy gap $\Delta_{21}$ is directly proportional to the inter-chain coupling, $\Delta_{21} = 2 J^{\text{eff}}_{\text{in}}$, allowing the effective coupling strength to be easily extrapolated from DMRG simulations~\cite{supplementary}.

 \begin{figure}[t]
 \includegraphics[width=8cm]{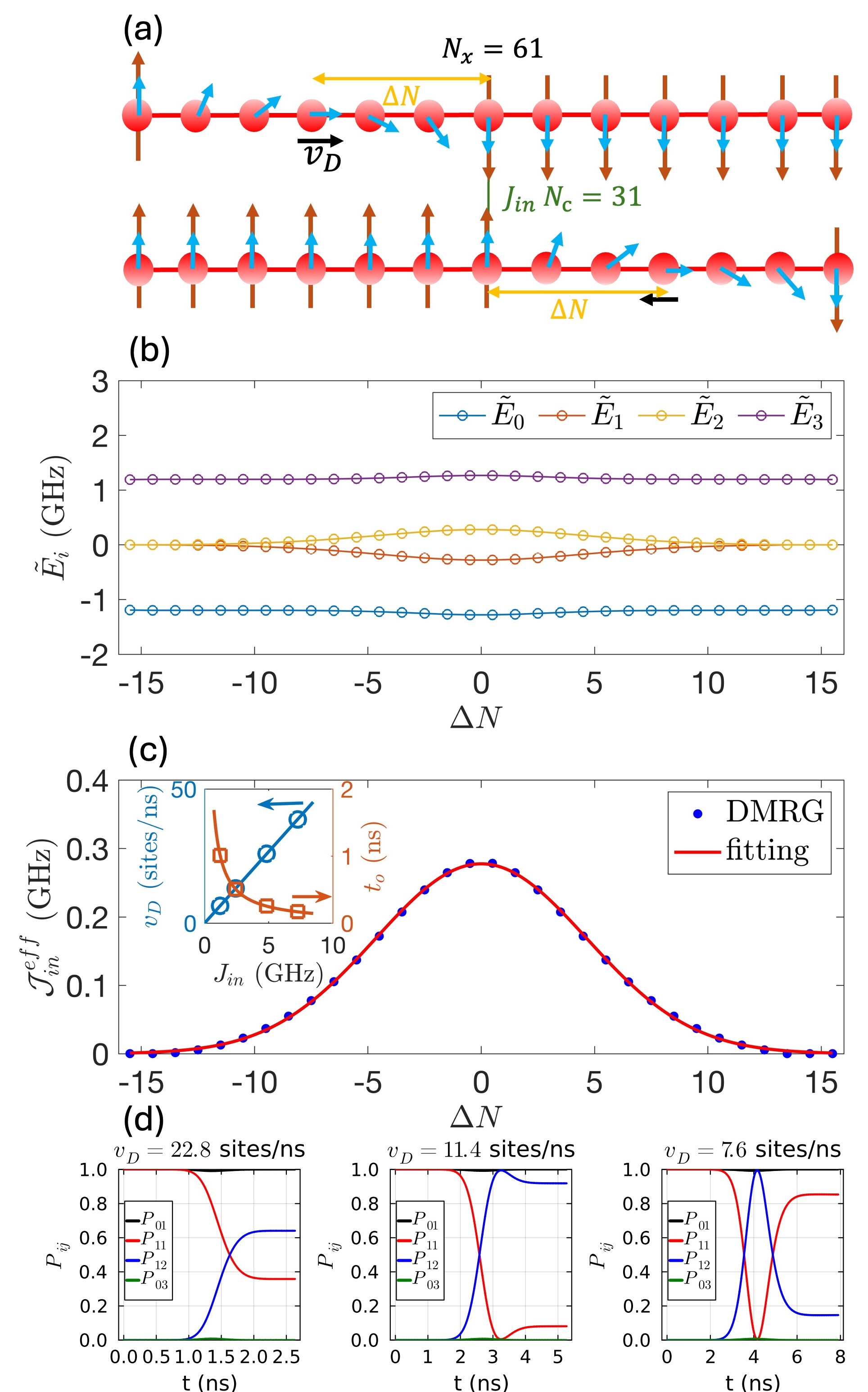}
 \caption{\label{fig:4} (a) Schematic illustration of DMRG simulation for two moving DW qubits. (b) Energy levels of two-qubit computational space for two moving DW qubits with displacement of DW center $\Delta N$. (c) Effective two-qubit coupling $\mathcal{J}^{\text{eff}}_{\text{in}} = \Delta_{21}/2$ as function of displacement of DW center and Gaussian profile with fitting parameters $J^{\text{eff}}_{\text{in}}$= 278 MHz and $N_D=6.58$. Single-site coupling strength is $J_{\text{in}}= 2.42$ GHz. The inset shows the required DW velocity $v_D$ and operation time $t_{\text{o}}$ as function of  single-site coupling strength. Parameters for each DW qubit are the same as in Fig.~\ref{fig:1}. (d) Transition probabilities $P_{ij}= |\braket{ \psi_{i}(t) | \psi_{j} }|^2$ for state $E_{0,1}$ under real-time evolution simulations of two mobile DW qubits with inter-chain coupling $\tilde{J}_{\text{in}} = 1$ $\mu$eV. The total simulation times are $t_{\text{tot}}=2.63,$ $5.26$, $7.89$ ns, corresponding to DW velocity of $v_{D}$= $22.8,$ $11.4$, $7.6$ sites/ns, respectively. The details of real-time simulation is shown in~\cite{supplementary}.} 
 \end{figure}
 
\section{V. Effective coupling between moving DWs}
Two-qubit gates can be achieved by shuttling two DWs toward each other via the spin-orbit torque~\cite{shaoRoadmapSpinOrbit2021, supplementary}. To investigate this process, we study the two-qubit interaction between DWs displaced by $2\Delta N$, as shown in Fig.~\ref{fig:4}(a).
Considering a spin ladder with a length greater than the domain wall width~($N_x=61$, $N=30$), the position of DWs is shifted by adjusting the pinning field $h_z$. With the inter-chain coupling only at the center of the chains~($N_c=31$), the energy levels of the two-qubit subspace are computed from DMRG simulation as a function of $\Delta N$ (see Fig.~\ref{fig:4}(b)).
When the DW centers are far apart with $\Delta N \gg N$, the hybridization gap~$\Delta_{21}$ vanishes. On the other hand, $\Delta_{21}$ reaches the maximum when the DW centers coincide with $\Delta N=0$.  As shown in Fig.~\ref{fig:4}(c),  we find that the effective coupling strength between the two moving DWs, as a function of DW displacement $\Delta N$, is well-fitted by a Gaussian profile:
\begin{align}
\mathcal{J}^{\text{eff}}_{\text{in}} (\Delta N) = J^{\text{eff}}_{\text{in}} \exp\left[ - (\Delta N/N_D)^2 \right] ,
\label{eq:7}
\end{align}
where $J^{\text{eff}}_{\text{in}}$ represents the maximum coupling  when the two DWs coincide, and $N_D$ defines the characteristic range over which the mobile DWs interact. The fitted value of $N_D$ is approximately $6.66$ sites, which is comparable to the DW width. Therefore, through this quantum simulation, we demonstrate that the interaction between DWs remains significant when their separation is less than the DW width, beyond which the interaction rapidly decays.

This position-dependent effective interaction  can be further utilized to implement two-qubit gate operations. To this end, we consider two DWs moving toward the coupling site at a velocity $v_D = \Delta N/t$, where $v_D$ represents the DW velocity in units of site/ns. 
In the interaction picture, the two-qubit interacting Hamiltonian reads
$\tilde{H}_{2} \approx  - \mathcal{J}^{\text{eff}}_{\text{in}}  (\Delta N) /2  \left( \sigma_x \otimes \sigma_x  +  \sigma_y \otimes \sigma_y \right)$, when the qubit frequency is much faster than the coupling rate between DW qubits.
Integrating over the Gaussian profile, the DW interaction yields a  two-qubit gate operation
\begin{align}
 U_2  
& =\exp\left[  \frac{i \pi^{3/2} J^{\text{eff}}_{\text{in}} N_D }{  v_D}   \left( \sigma_x \otimes \sigma_x  +  \sigma_y \otimes \sigma_y \right) \right].
\label{eq:9}
\end{align}
For a DW with velocity  $v_D= 4 \sqrt{\pi} J^{\text{eff}}_{\text{in}} N_D$, the unitary operation $U_2$ acts as an XY-gate, which is equivalent to a controlled-NOT gate up to single-qubit rotations. Together with single qubit rotations, we then have a set of universal quantum gates.   From the Gaussian fitting in Fig.~\ref{fig:4}(c), we extract the amplitude of the effective coupling as $J^{\text{eff}}_{\text{in}} = 278$~MHz and the interaction range as $N_D = 6.58$, corresponding to a DW velocity of $v_D = 13$~sites/ns and an ultrafast two-qubit gate operation time of approximately $1$~ns. Importantly, the interaction range between two DWs, characterized by $N_D$, is independent of the single-site coupling strength $J_{\text{in}}$. Consequently, the required DW velocity exhibits a linear dependence on $J_{\text{in}}$, while the operation time $t_{\text{o}}$ is inversely proportional to $J_{\text{in}}$, as shown in Fig.~\ref{fig:4}(c). As the inter-chain coupling can be tuned, for instance, by adjusting the distance between spin chains, our system offers flexibility in controlling the operational time of the two-qubit gate.

In Fig.~\ref{fig:4}(d), we show the real-time evolution of a two-qubit gate operation for two moving DWs, where we confirm the XY-gate operation at a small inter-chain coupling strength of \( \tilde{J}_{\text{in}} = 1 \) $\mu$eV. To achieve an i-swap gate, the DW velocity is required to be approximately $v_D\sim11.4$ sites/ns. Note that \( \tilde{J}_{\text{in}} \) in the real-time simulations must be scaled by a factor of 10 for comparison with the single-site coupling \( J_{\text{in}} \)~\cite{supplementary}.

Compared to other quantum computing platforms~\cite{supplementary}, DW qubits excel in their compact size, which is comparable to that of spin qubits ($\sim 100\,\mathrm{nm}^2$). This highlights their potential compatibility and integrability with spin-based architectures. Furthermore, DW qubits exhibit a gate rate of up to $1\,\mathrm{GHz}$, surpassing that of all existing quantum platforms, underscoring their promise for high-speed quantum processing. These advantages collectively position DW qubits as a  scalable and high-speed candidate in the landscape of quantum computing devices.

\section{VI. Conclusion}
We have demonstrated the feasibility of utilizing DWs as qubits for universal quantum computation in a fully quantum framework. Through DMRG simulations, we established the existence of a well-defined low-energy subspace formed by DW chirality states, enabling robust quantum information encoding. The tunability of qubit splittings via magnetic fields, characterized by anisotropic effective $g$-factors, offers a practical way for implementing single-qubit gates. Moreover, we constructed an effective Hamiltonian for two interacting DW qubits, revealing the mechanism for generating entanglement and implementing two-qubit gates.
Employing DW qubits as stationary qubits and flying qubits, a scalable two-dimensional quantum computing platform could be realized~\cite{supplementary}. In addition, it could facilitate a long-range quantum transmission between spin qubits~\cite{zouTopologicalSpinTextures2024}.
Our work marks a pivotal step from semiclassical treatment to a fully quantum framework of nanoscale topological spin textures. By bridging this critical gap, we establish a solid foundation for utilizing topological spin textures in universal quantum computation.

\begin{acknowledgments}
This work was supported by the Georg H. Endress Foundation and by the Swiss National Science Foundation, NCCR SPIN (grant number 51NF40-180604). T.~H. is supported by JSPS KAKENHI Grant Number JP23K13064. The numerical simulations were carried out on the HOKUSAI supercomputing system at RIKEN
(Project ID No. RB230055). The codes of DMRG simulations are written based on the ITensor Library~\cite{ITensor}.
\end{acknowledgments}

\bibliography{Domain_wall_Qubit_ArXiv}

\clearpage
%
%

\title{Supplemental Material: \\ ``Density Matrix Renormalization Group Study of Domain Wall Qubits"}

\author{Guanxiong Qu}
\affiliation{RIKEN, Center for Emergent Matter Science (CEMS), Wako-shi, Saitama 351-0198, Japan} 
\author{Ji Zou}
\affiliation{Department of Physics, University of Basel, Klingelbergstr.~82, 4056 Basel, Switzerland}
\author{Daniel Loss}
\affiliation{RIKEN, Center for Emergent Matter Science (CEMS), Wako-shi, Saitama 351-0198, Japan} 
\affiliation{Department of Physics, University of Basel, Klingelbergstr.~82, 4056 Basel, Switzerland}
\affiliation{RIKEN, Center for Quantum Computing (RQC), Wako-shi, Saitama 351-0198, Japan}
\author{Tomoki Hirosawa}
\affiliation{College of Science and Engineering, Aoyama Gakuin University, Japan}

\maketitle
\begin{widetext}

\begin{center}
    {\large \textbf{Supplemental Material: \\``Density Matrix Renormalization Group Study of Domain Wall Qubits"}}\\[10pt]
    { Guanxiong Qu$^{1}$, Ji Zou$^{2}$, Daniel Loss$^{1,2,3}$, Tomoki Hirosawa$^{4}$}\\
    { \textit{1. RIKEN, Center for Emergent Matter Science (CEMS), Wako-shi, Saitama 351-0198, Japan}}\\
    { \textit{2. Department of Physics, University of Basel, Klingelbergstr. 82, 4056 Basel, Switzerland}}\\
     {\textit{3. RIKEN, Center for Quantum Computing (RQC), Wako-shi, Saitama 351-0198, Japan}}\\
    { \textit{4. Department of Physical Science, Aoyama Gakuin University, Kanagawa 252-5258, Japan}}\\
    {\today}
\end{center}

\section{DMRG simulations on a single DW qubit}\label{Sup:A}

\subsection{ Effect of in-plane anisotropy  $K_y$}
In Figure~\ref{fig:S1}, we illustrate the tuning of in-plane anisotropy ($K_y$) for both odd ($N=31$) and even ($N=30$) single spin chains in the absence of an external magnetic field$(h_x=h_y=0)$. For XXZ spin chains ($K_y=0$), the ground state of an $N$-even spin chain is non-degenerate, whereas the ground state of an odd spin chain is doubly degenerate.This distinction arises because the boundary interaction $h_z$  lifts the Kramers degeneracy in $N$-even chains but not in $N$-odd chains~\cite{schilling_quantum_1977}. However, with the introduction of $K_y$, the ground state $E_0$ and the first excited state $E_1$  of the $N$-even spin chain also become degenerate, effectively reducing the finite-size parity effect for sufficiently large in-plane anisotropy ($K_y >0.04$ meV), as shown in Figure~\ref{fig:S1} (c). Furthermore, we observe that the in-plane anisotropy $K_y$ also increases the energy gap $\Delta_{21}$ between the first degenerate subspace (computational qubit space) and the second degenerate subspace.

\begin{figure}[h]
    \centering
    \includegraphics[width=0.8\textwidth]{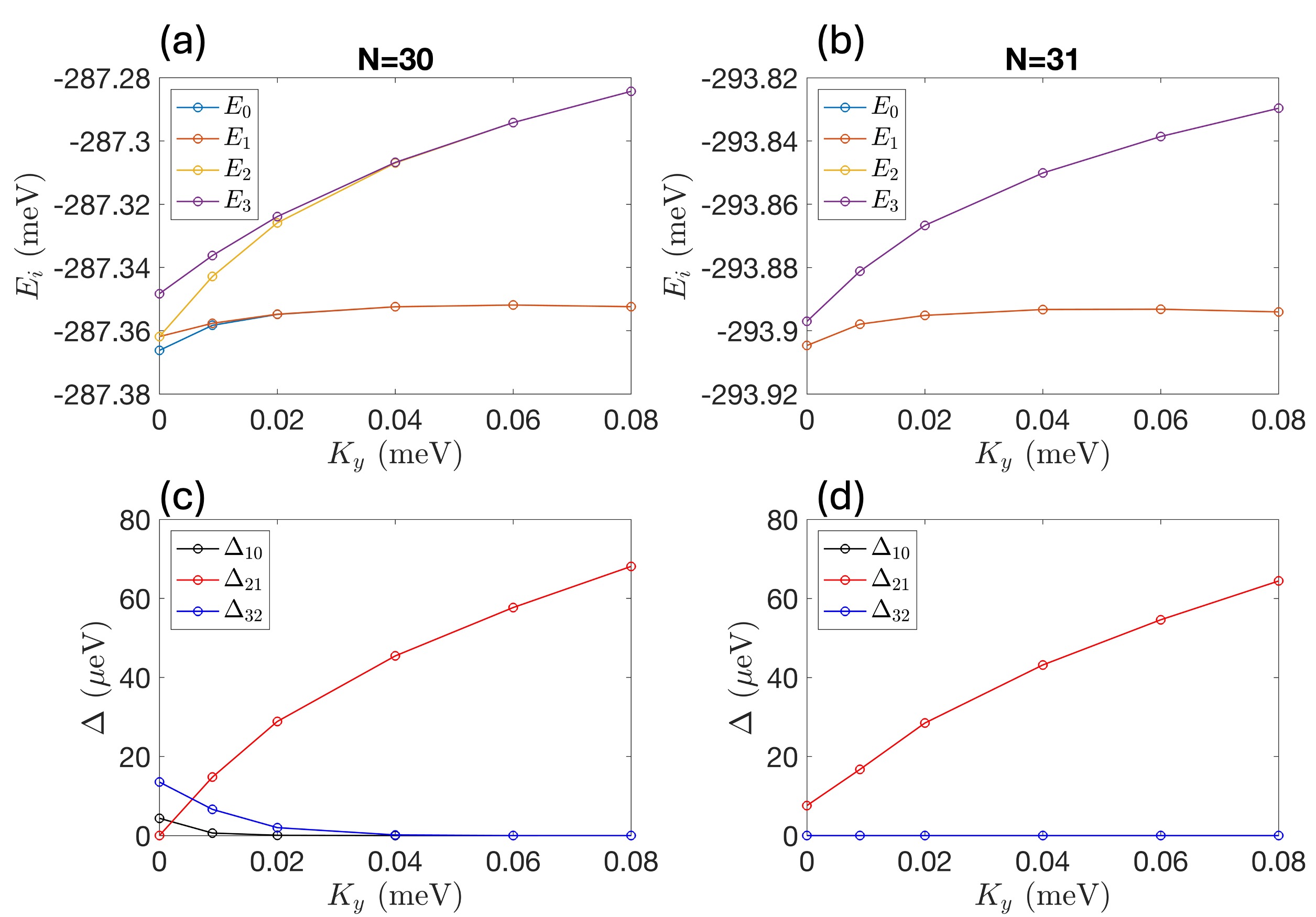}
    \caption{Energy spectra of (a,b) the four lowest-lying states $E_i$ and (c,d) the energy gaps $\Delta_{ij} \equiv E_i - E_j$ as tuned by in-plane anisotropy $K_y$ for even $N = 30$ (left panel) and odd $N = 31$ (right panel) spin chains. The parameters used in the DMRG simulations are $J = 25.85$ meV, $K_z = 0.26$ meV, and $\mu_B h_z = -100$ meV. The criterion of energy convergence is at least $10^{-10}$ meV. }
    \label{fig:S1}
\end{figure}

\subsection{ Effect of external magnetic fields  $h_x,h_y$}

In Figure~\ref{fig:S2}, we demonstrate the effect of external magnetic fields on the qubit splitting $\Delta_{10}$ and the gap between the qubit space and higher energy levels $\Delta_{21}$ of a single DW qubit. The gap $\Delta_{21}$ decreases linearly with increasing $h_x$ until $\Delta_{21}$ closes, while the qubit splitting $\Delta_{10}$ increases linearly with $h_x$, independent of the magnitude of in-plane anisotropy $K_y$ [see Figure~\ref{fig:S2}(a,c)]. In contrast, the $h_y$ field causes an exponential increase in $\Delta_{10}$ beyond certain thresholds~\cite{zou_quantum_2023}, while $\Delta_{21}$ is only minimally affected by the $h_y$ field. These thresholds for qubit splitting opening with the $h_y$ field grow as the in-plane anisotropy $K_y$ increases.

To construct the DW qubit, a bias $h_y$ field is preferred for opening the qubit splitting $\Delta_{10}$, as it exponentially increases $\Delta_{10}$ while keeping $\Delta_{21}$ nearly unchanged. A small $h_x$ field can be employed as a tuning field for the DW qubit due to its linear relationship with the qubit splitting $\Delta_{10} \propto h_x$. In the main text, we choose $K_y = 0.1$ meV and $h_y = 0.9$ T, where the qubit splitting is $\Delta_{10} \sim 4$ $\mu$eV and the gap between the qubit space and higher levels is $\Delta_{21} \sim 40$ $\mu$eV. Thus, the qubit subspace is separated by a gap from the higher energy levels.
\begin{figure}[h]
    \centering
    \includegraphics[width=0.8\textwidth]{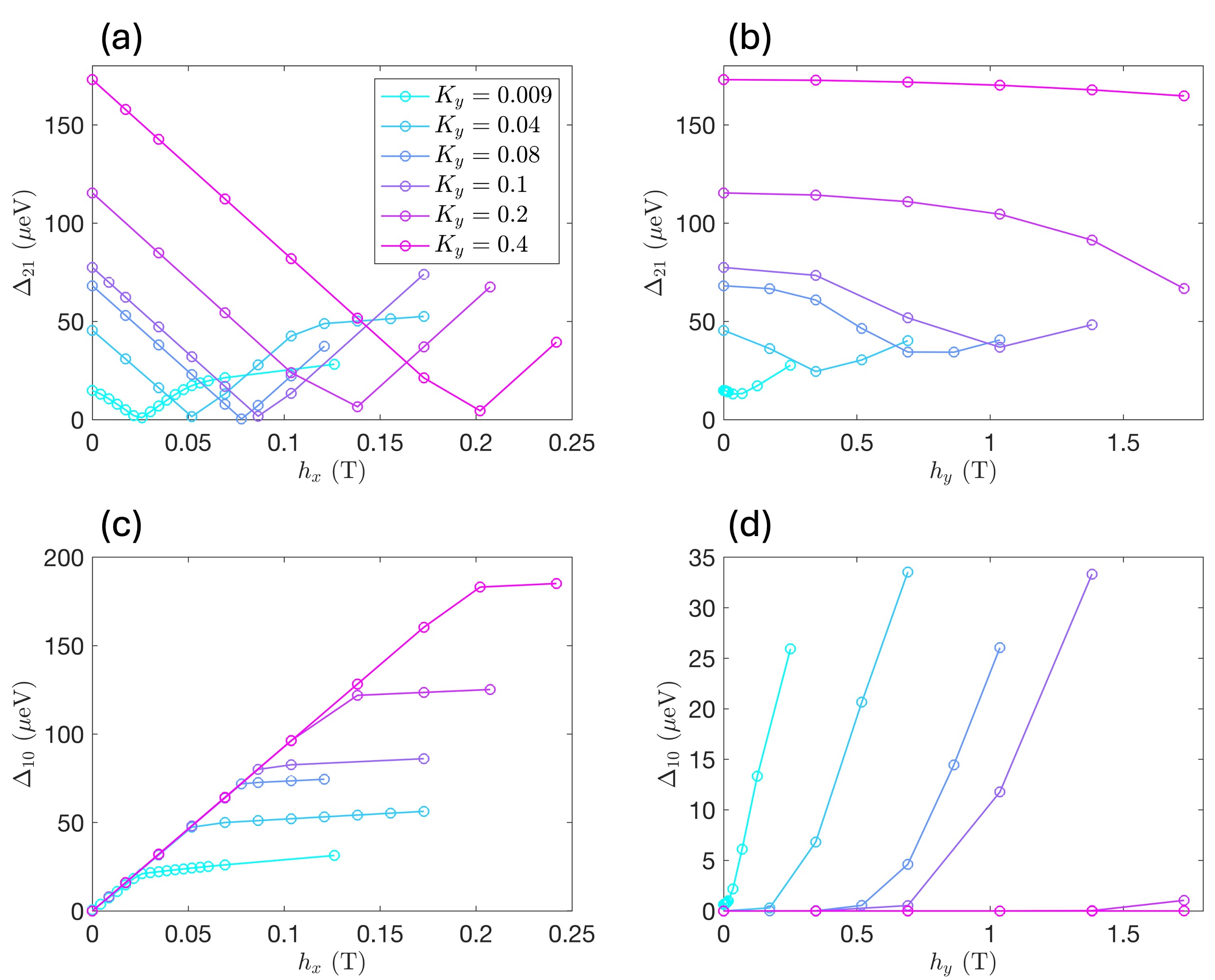}
    \caption{(a,b) qubit splitting $\Delta_{10}$ and (c,d) the gap between the qubit space and higher energy levels $\Delta_{21}$, tuned by external fields $h_x$ and $h_y$ under various in-plane anisotropies $K_y$. The parameters used in the DMRG simulations are $N = 30$, $J = 25.85$ meV, $K_z = 0.26$ meV, and $\mu_B h_z = -100$ meV.  The criterion of energy convergence is at least $10^{-10}$ meV.}
    \label{fig:S2}
\end{figure}

\subsection{Comparison between $N$-even and $N$-odd spin chains}
\begin{figure}[h]
    \centering
    \includegraphics[width=0.8\textwidth]{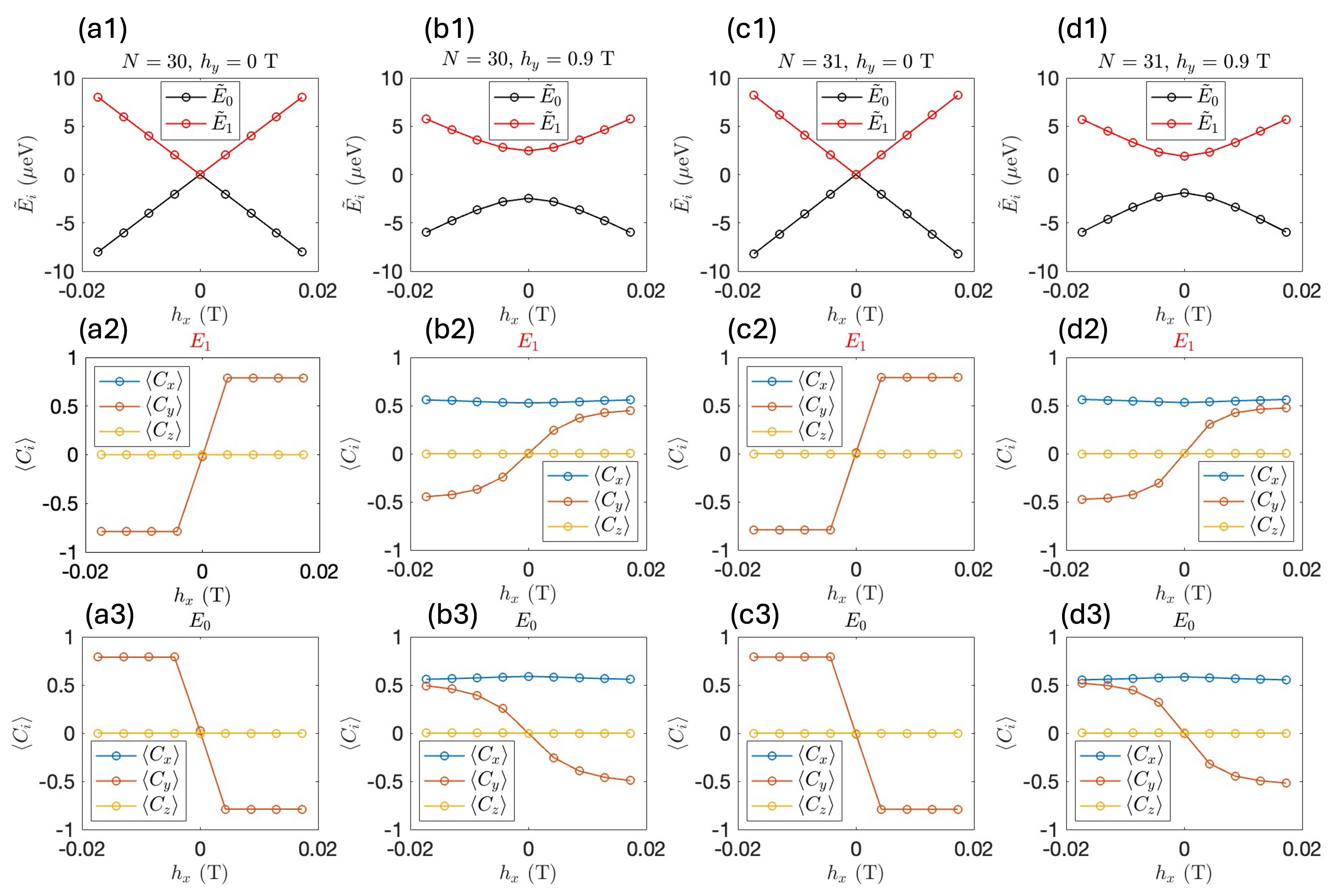}
    \caption{DMRG simulation data  of (1) DW qubits energy spectra  and  (2,3) DW chiralities of each qubit states with tuning of external magnetic fields  $h_x$ for (a,b) even number spin chains and (c,d) odd number spin chain. The parameters used in the DMRG simulations are $N = 30,31$, $J = 25.85$ meV, $K_z = 0.26$ meV, $K_y = 0.1$ meV, and $\mu_B h_z = -100$ meV.  The criterion of energy convergence is at least $10^{-10}$ meV.}
    \label{fig:S8a}
\end{figure}

In Figure~\ref{fig:S8a}, we compare spin-$1/2$ chains with an even and an odd number $N$ of sites. For a sufficiently large $K_y = 0.26$~meV, no significant difference is observed between the even and odd number chains, and both require a substantial bias field ($h_y \sim 0.9$~T) to open the qubit gap.

\section{DMRG simulations on coupled DW qubits}\label{Sup:B}

\subsection{Uniform  inter-chain coupling}
We consider two DW spin chains uniformly coupled at each site [see Figure~\ref{fig:S3} (a)]. The inter-chain coupling Hamiltonian reads
\begin{eqnarray}
     H_{inter}&= - J_{in} \sum^{N}_{i=1}  \bm{S}_{i,1} \cdot \bm{S}_{i,2},
         \label{eq:S1}
\end{eqnarray}
where $J_{in}>0$ denotes ferromagnetic coupling constant uniformly across the spin chains. 

\begin{figure}[h]
    \centering
    \includegraphics[width=0.8\textwidth]{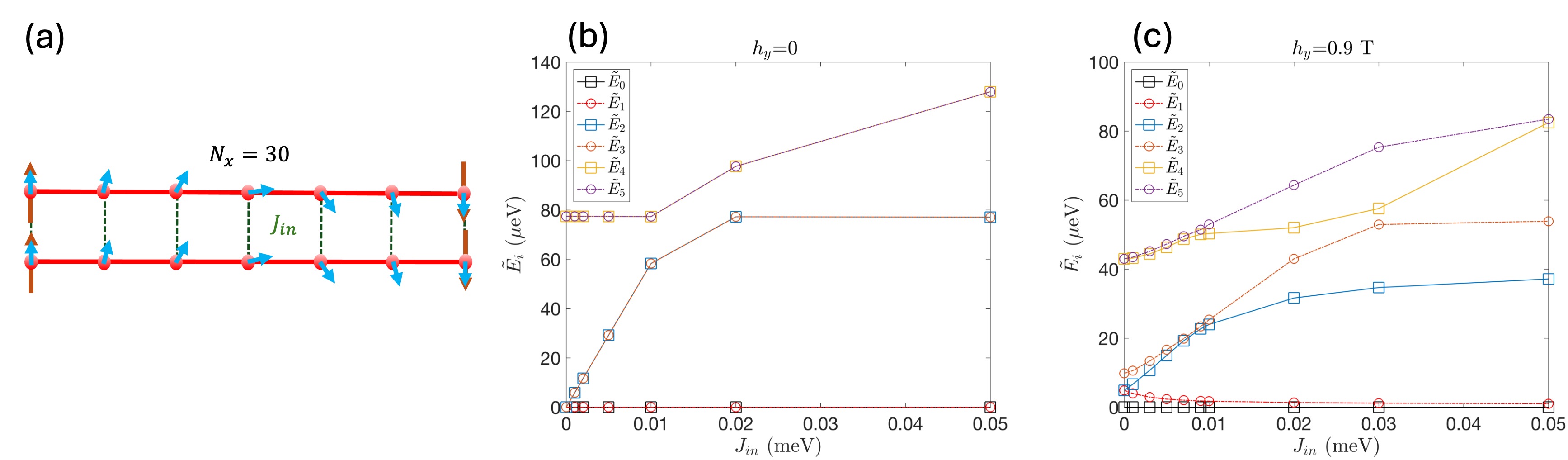}
    \caption{(a) Schematic plot of two uniformly coupled DW qubits. Energy spectra of spin ladders with inter-chain coupling $J_{in}$ at (a) $h_y=0$~T and (b) $h_y=0.9$~T. The parameters used in the DMRG simulations are $N = 30$, $J = 25.85$ meV, $K_z = 0.26$ meV, $K_y = 0.1$ meV, and $\mu_B h_z = -100$ meV.  
  The criterion of energy convergence is at least $10^{-10}$ meV.   }
    \label{fig:S3}
\end{figure}
Figure~\ref{fig:S3} (b,c) presents the energy spectra of two coupled spin chains against the inter-chain coupling strength $J_{in}$. Note that the DW qubit splitting closes on a single spin chain without applying the $h_y$ field. On coupled spin chains, the DW qubit splitting on each chain remains closed without $h_y$ fields, resulting in four degenerate states at $J_{in}=0$. As the inter-chain coupling is turned on, the four degenerate states split into doubly degenerate ``bonding" and ``anti-bonding" states between the two spin chains, as shown in Fig.~\ref{fig:S3}(b). Notably, the gap between the ``bonding" and ``anti-bonding" states saturates at a very small inter-chain coupling strength $J_{in} \sim 0.02$ meV, compared to the intra-chain coupling $J = 25.85$ meV.

When the $h_y$ field is applied, the DW qubit splitting emerges in each spin chain. The two levels of the DW qubit are labeled as $\ket{0} = (\ket{\circlearrowright}+\ket{\circlearrowleft})/\sqrt{2}$ and $\ket{1} = (\ket{\circlearrowright}-\ket{\circlearrowleft})/\sqrt{2}$. In the absence of inter-chain coupling ($J_{\text{in}}=0$), two identical spin chains remain decoupled, where the coupled DW qubit subspace exhibits three energy levels with corresponding eigenstates: $\ket{00}$, $\ket{01}$/$\ket{10}$, and $\ket{11}$, where $\ket{01}$ and $\ket{10}$  are the doubly degenerate states. The introduction of inter-chain coupling lifts the degeneracy of the $\ket{01}$ and $\ket{10}$ states, resulting in two ``bonding" and ``anti-bonding" pairs of states in the coupled DW qubit subspace, as illustrated in Fig.~\ref{fig:S3}(c). The gap between the ``bonding" and ``anti-bonding" pairs nearly saturates at $J_{\text{in}} \sim 0.02$ meV, similar to the case without the $h_y$ field. In the following, we focus on the regime where the two spin chains are weakly coupled ($J_{\text{in}} < 5$ $\mu$eV), and the ``bonding" and ``anti-bonding" pairs are moderately gapped. 

\begin{figure}[h]
    \centering
    \includegraphics[width=0.8\textwidth]{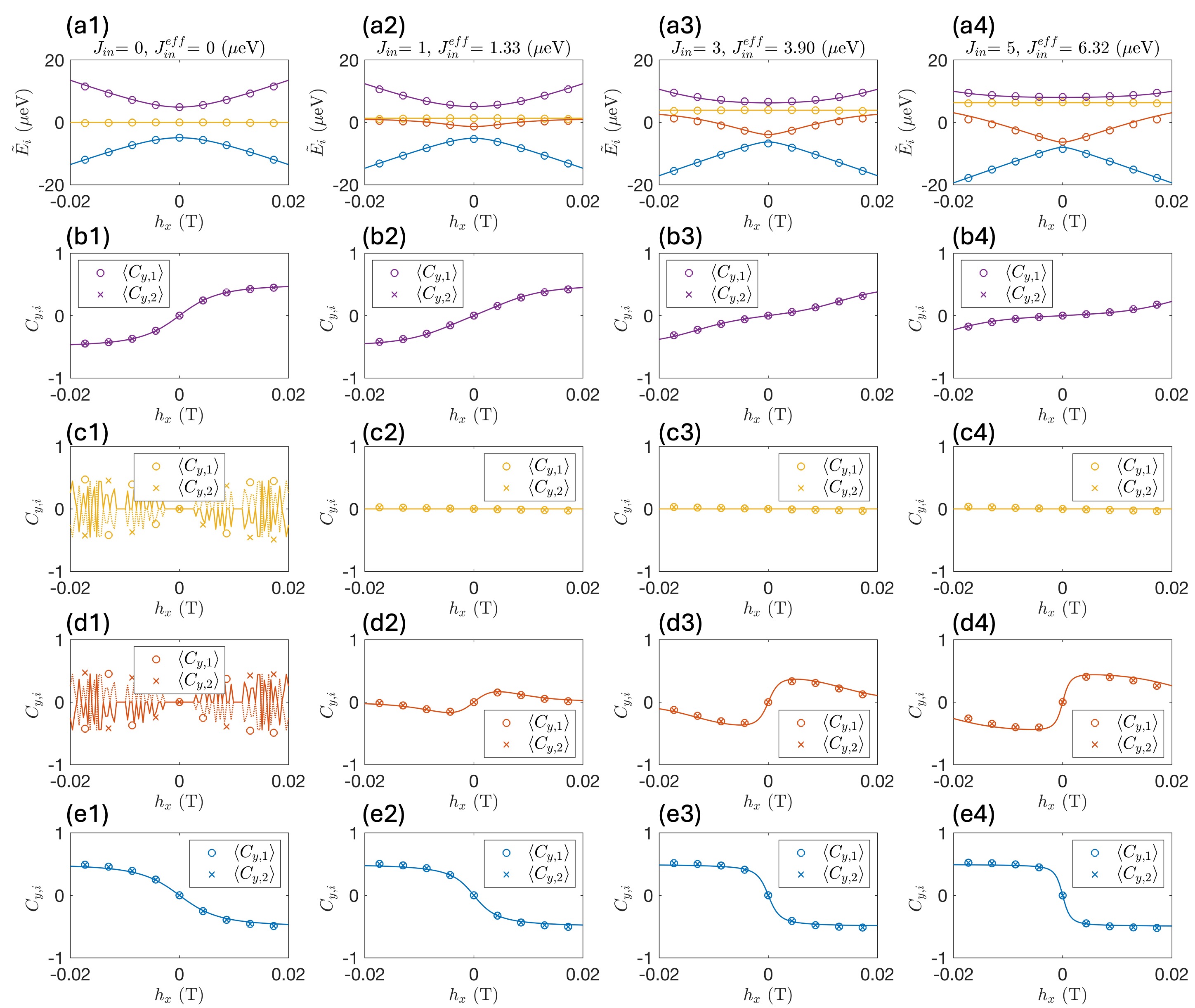}
    \caption{DMRG simulation data (circles and crosses) and effective model data (lines) of (a) energy spectra and (b-e) DW chirality $C_{y,i}$ of each spin chain ($i=1,2$) with various inter-chain coupling strength $J_{in}=0,1,3,5$ $\mu$eV and corresponding effective coupling strength $J^{eff}_{in}=0, 1.33, 3.90, 6.32$ $\mu$eV.  The parameters used in the DMRG simulations are $N = 30$, $J = 25.85$ meV, $K_z = 0.26$ meV, $K_y = 0.1$ meV, $h_y =0.9$ T, and $\mu_B h_z = -100$ meV.  The criterion of energy convergence is at least $10^{-10}$ meV. }
    \label{fig:S4}
\end{figure}

Figure~\ref{fig:S4} (a) shows the energy spectra of DW qubits under varying $h_x$ fields for different inter-chain coupling strengths, $J_{\text{in}} = 0, 1, 3, 5 \, \mu\text{eV}$. As the inter-chain coupling strength $J_{\text{in}}$ increases, the energy gap between the $\ket{01}$ and $\ket{10}$ states widens, indicating an interaction between the two DW qubits. We also present the DW chiralities $C_{y,i}$ of each spin chain ($i=1,2$) for each energy level in the DW qubit subspace, as shown in Figure~\ref{fig:S4} (b-e). Remarkably, the chiralities $C_{y,i}$ of the two spin chains always coincide, suggesting that the spin chains exhibit coherent interaction under ferromagnetic inter-chain coupling. For decoupled spin chains ($J_{\text{in}} = 0$), we find that the two non-degenerate states $E_0$ and $E_3$ have opposite DW chiralities, while the expectation value of DW chirality for the doubly degenerate states is zero [see Fig.~\ref{fig:S4} (c1,d1)]. As $J_{\text{in}}$ increases, the DW chiralities $C_{y,i}$ for the $E_3$-states decrease, while $C_{y,i}$ for the $E_1$-states increase.

\subsection{Effective Hamiltonian of coupled domain wall qubit}
We recall the effective Hamiltonian of coupled DW qubits shown in the main text [Eq.~(9)]:
\begin{align} 
    H^{\text{eff}}_{2} &= H^{\text{eff}}_1 \otimes I_2 +  I_2 \otimes H^{\text{eff}}_1  - J^{\text{eff}}_{\text{in}} \sigma_x \otimes \sigma_x \notag \\
    &= \left(\begin{array}{cccc}
     \Delta_{10} + 2 g_y \Delta h_y & - g_x h_x & -g_x h_x & -  J^{\text{eff}}_{\text{in}} \\     
      -g_x h_x & 0 & -  J^{\text{eff}}_{\text{in}} & -g_x h_x \\ 
    -g_x h_x & -  J^{\text{eff}}_{\text{in}} & 0 &  -g_x h_x \\
     -  J^{\text{eff}}_{\text{in}} & -g_x h_x & -g_x h_x &  - \Delta_{10} -  2 g_y \Delta h_y
    \end{array}\right).
    \label{eq:S2}
\end{align}
Without driving fields ($h_x=0$, $\Delta h_y=0$), the eigenenergies of the coupled DW qubits are trivial: $\pm J^{\text{eff}}_{\text{in}}$, $\pm \sqrt{ \Delta_{10}^2/4 + {J^{\text{eff}}_{\text{in}}}^2}$. For weak coupling ($J^{\text{eff}}_{\text{in}} < \Delta_{10}/2$), the energy gap between the $\ket{01}$ and $\ket{10}$ states is proportional to the effective inter-chain coupling, $\Delta_{21} \equiv E_2 - E_1 = 2 J^{\text{eff}}_{\text{in}}$. Thus, the effective inter-chain coupling strength $J^{\text{eff}}_{\text{in}}$ in Eq.~\eqref{eq:S2} can be easily extrapolated from the gap size $\Delta_{21}$ in DMRG simulations.  The DW chirality operator for each spin chain in the effective Hamiltonian is defined as $\hat{C}_{y,1}=  \frac{1}{2} \sigma_x \otimes I_2$,  $\hat{C}_{y,2}= \frac{1}{2} I_2 \otimes\sigma_x $. Figure~\ref{fig:S4} shows the energy spectra and corresponding DW chiralities of each energy level with $J^{\text{eff}}_{\text{in}}$ extrapolated from DMRG simulations. The data from the effective Hamiltonian calculation are consistent with DMRG simulations.

\subsection{Single-site inter-chain coupling}
\begin{figure}[h]
    \centering
    \includegraphics[width=0.8\textwidth]{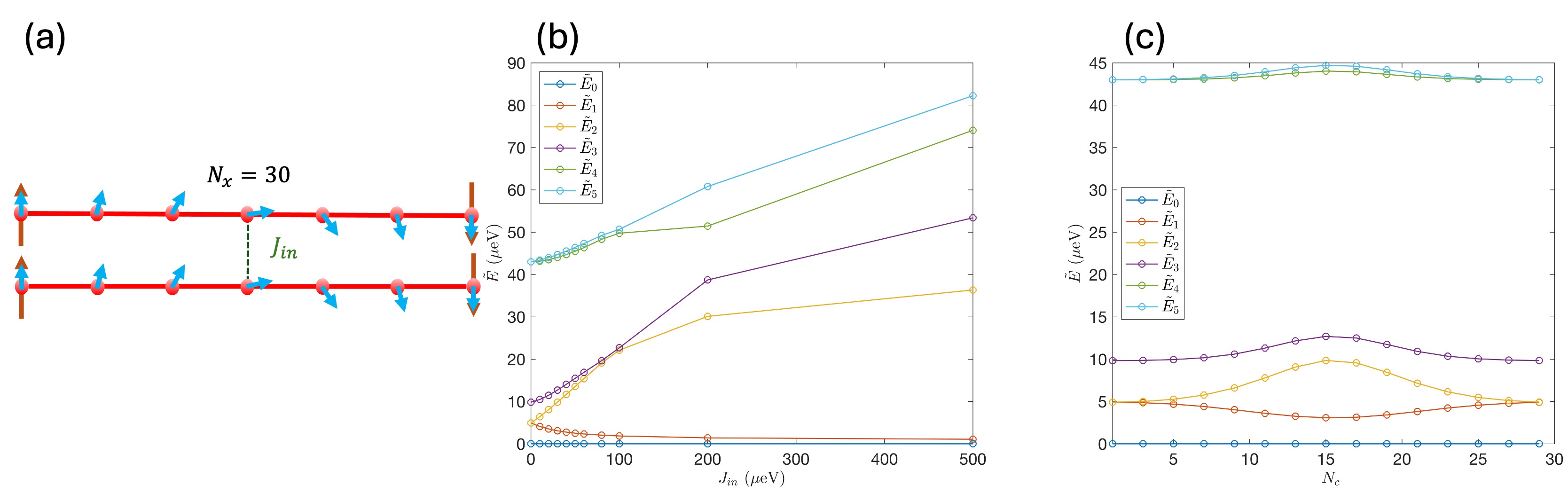}
    \caption{(a) Schematic plot of single-site coupled DW qubits. Energy spectra of coupled DW qubits with (b) fixed coupling site ($N_c=15$) and (c) fixed coupling strength $J_{in}=30$ $\mu$eV. The parameters used in the DMRG simulations are $N = 30$, $J = 25.85$ meV, $K_z = 0.26$ meV, $K_y = 0.1$ meV, and $\mu_B h_z = -100$ meV.  The criterion of energy convergence is at least $10^{-10}$ meV.
    }
    \label{fig:S6}
\end{figure}
For single-site coupled spin chains [see Fig.~\ref{fig:S6} (a)], the inter-chain coupling Hamiltonian is 
\begin{align}
H_{inter}= - J_{in}   \bm{S}_{N_c,1} \cdot \bm{S}_{N_c,2},
\label{eq:S3}
\end{align}
where $N_c$ is the site at which the two spin chains are  exchange coupled.

Figure~\ref{fig:S6}(b) shows energy spectra of the single-site coupled spin chains against the inter-chain coupling strength $J_{in}$. The energy spectra show the same profile as uniformly coupled spin chains, Fig.~\ref{fig:S3}(c), while the magnitude of coupling strength required to open the same gap $\Delta_{12}$ is one order of magnitude larger. 
The energy spectra at different coupling sites is shown in Fig.~\ref{fig:S6}(c). The interaction between two DW qubits reaches its maximum at the center of DWs and is nearly zero at the boundaries.
We also confirm that the two-qubit subspace with the single-site coupling and uniform coupling is equivalent with small $J_{in}$, see Fig.~\ref{fig:S7}.

\begin{figure}[h]
    \centering
    \includegraphics[width=0.8\textwidth]{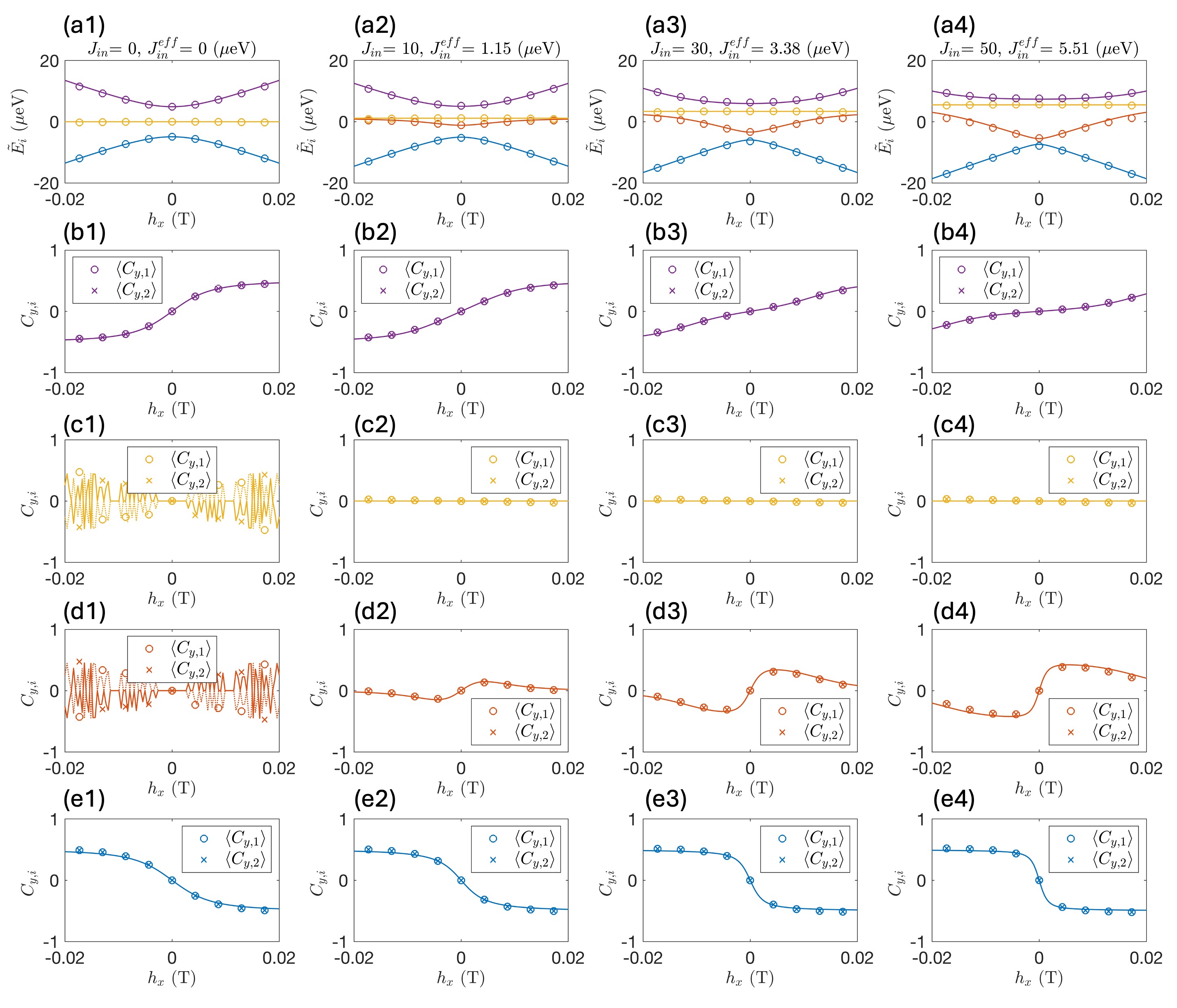}
    \caption{DMRG simulation data (circles and crosses) and effective model data (lines) of (a) energy spectra and (b-e) DW chirality $C_{y,i}$ of each spin chain ($i=1,2$) with various single-site inter-chain coupling strength $J_{in}=0,10,30,50$ $\mu$eV and corresponding effective coupling strength  $J^{eff}_{in}=0, 1.15, 3.38, 5.51$ $\mu$eV.  The parameters used in the DMRG simulations are $N = 30$, $J = 25.85$ meV, $K_z = 0.26$ meV, $K_y = 0.1$ meV, $h_y =0.9$ T, and $\mu_B h_z = -100$ meV.  The criterion of energy convergence is at least $10^{-10}$ meV.
    }
    \label{fig:S7}
\end{figure}

\subsection{Two mobile domain walls}
Figure~\ref{fig:S8}(a) displays the energy spectra of two mobile DWs as a function of the displacement between their centers, $\Delta N$. As the single-site coupling strength $J_{in}$ increases, the energy gap $\Delta_{21}$ widens, indicating stronger coupling as the two DWs move closer to each other. The effective coupling strengths $\mathcal{J}^{eff}_{in}$, extrapolated from DMRG simulations, are fitted with Gaussian profiles of $\Delta N$, see Fig.~\ref{fig:S8}(b). The amplitude of these Gaussian profiles scales with $J_{in}$, while the coupling regime between the two mobile DWs, $N_D$, remains constant.
\begin{figure}[h]
    \centering
    \includegraphics[width=0.8\textwidth]{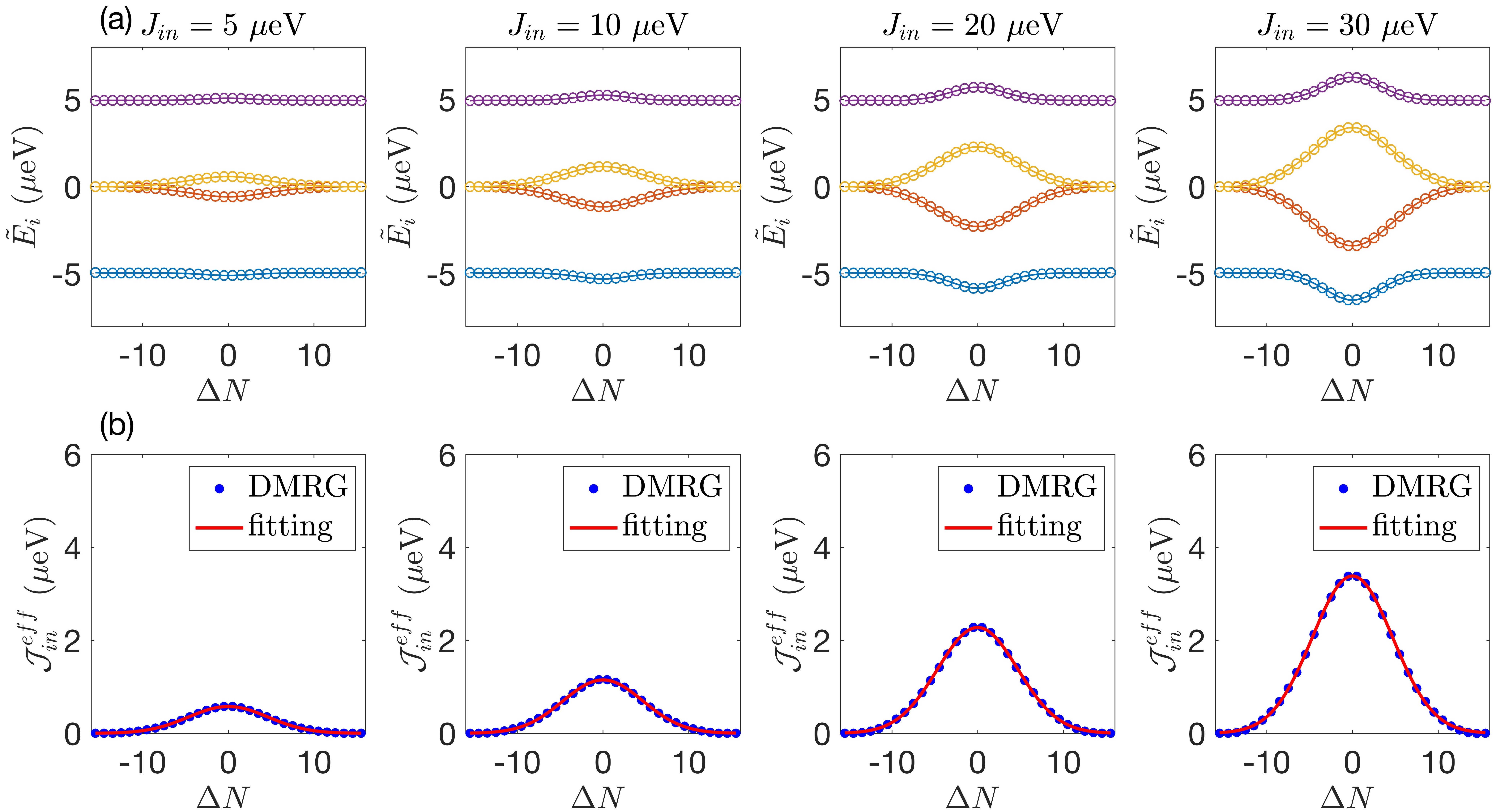}
    \caption{ DMRG simulation data of (a) energy spectra and (b) effective coupling strength as functions of the displacement of DW centers, $\Delta N$, for various single-site coupling strengths $J_{in} = 5, 10, 20, 30 \, \mu \text{eV}$.  The parameters used in the DMRG simulations are $N_x = 61$, $N_c=31$, $N=30$, $J = 25.85$ meV, $K_z = 0.26$ meV, $K_y = 0.1$ meV, $h_y =0.9$ T, and $\mu_B h_z = -100$ meV.  The criterion of energy convergence is at least $10^{-9}$ meV. }
    \label{fig:S8}
\end{figure}

\section{Qubit gate operations}

\subsection{Single qubit gate operation}
For a single DW qubit, the Rabi-driving Hamiltonian reads:
\begin{equation}
    H^{\text{Rabi}}_1 = \frac{\hbar \omega_{q}}{2} \sigma_z -  g_x h_x \cos(\omega_x t) \sigma_x,
    \label{eq:9}
\end{equation}
where $\hbar \omega_q = \Delta_{10}$. In the rotating frame, ignoring higher-order harmonics, the Hamiltonian reads:
\begin{equation}
    \tilde{H}_{\text{RWA}} = \frac{\hbar \Delta \omega}{2} \sigma_z -  \frac{g_x h_x}{2} \sigma_x,
    \label{eq:10}
\end{equation}
where $\Delta \omega = \omega_{q} - \omega_x$. This yields the time-evolution operator:
\begin{equation}
    U(t) = \exp\left\{ - i\left( \frac{\Delta \omega}{2} \sigma_z -  \frac{g_x h_x}{2 \hbar} \sigma_x \right)t \right\}.
    \label{eq:11}
\end{equation}
which rotates the qubit state around the vector $\left(- \frac{g_x h_x}{2 \hbar}, 0, \frac{\Delta \omega}{2} \right)$.

\subsection{Real-time simulation on single DW qubit}
We perform real-time evolution on quantum spin chain to simulate single-qubit rotation under linear Rabi driving. The time-dependent part of the Hamiltonian is given by
\begin{align}
H_t (t) = h_x \cos(\omega_x t) \sum^N_{i=1} S^x_{i}.
\label{eq:18}
\end{align}
The time evolution is computed using the time-dependent variational principle (TDVP) with a Krylov subspace approach~\cite{PhysRevLett.107.070601}. For the effective model [Eq.~\eqref{eq:9}], the transition probability from $E_0$ to $E_1$ states under the resonance condition $\omega_x = \omega_q$ is
\begin{align}
P_{01,Floquet} =\sin^2 ( \pi g_x h_x  t ),
\label{eq:19}
\end{align}
where $g_x$ is in unit of GHz/T.

\begin{figure}[h]
    \centering
    \includegraphics[width=0.8\textwidth]{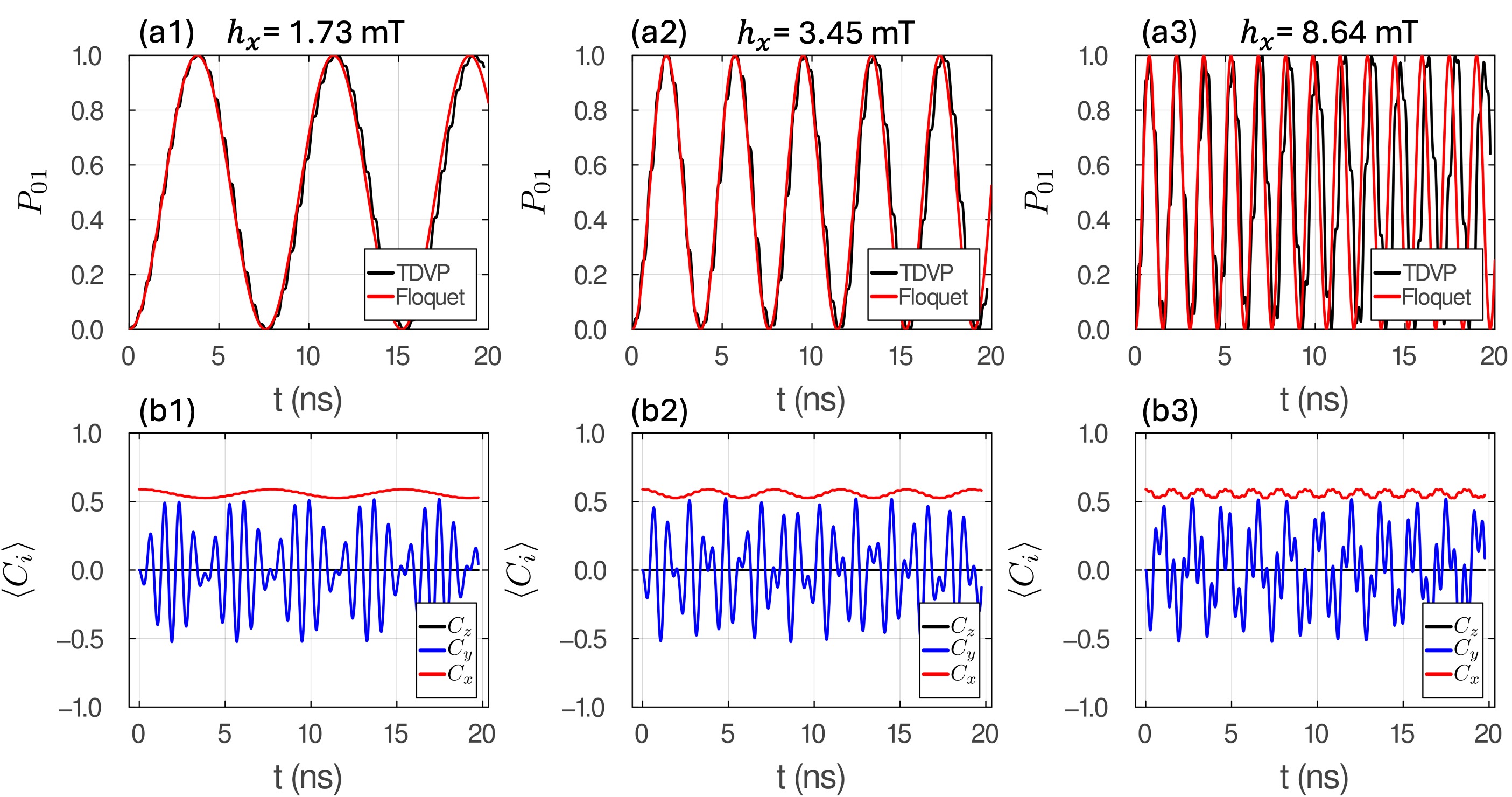}
    \caption{ (a) Transition probability $P_{01}$ and (b) DW chiralities $\braket{C_i}$ under real-time evolution for different field strengths of linear Rabi driving, with $h_x=1.73$, $3.45$, $8.64$ mT at the resonant condition $\omega_x=\omega_q$. The parameters of the single DW qubit are the same with DMRG simulations: $N=30$, $J = 25.85$ meV, $K_z = 0.26$ meV, $K_y = 0.1$ meV, $h_y =0.9$ T, and $\mu_B h_z = -100$ meV. The time step is $\tau=6.58\times 10^{-2}$ ps for all simulations.
     }
    \label{fig:S9}
\end{figure}

In Figure~\ref{fig:S9}, we present the time evolution of a single DW qubit under linear Rabi driving. The transition probability $ P_{01} = |\braket{\psi_0 (t) | \psi_1}|^2 $ exhibits Rabi oscillations with different frequencies, corresponding to varying field strengths $ h_x $ [see Figure~\ref{fig:S9}(a)]. The single qubit rotation can be achieved with high fidelity $F_{\text{single}}>99.9 \%$, which is estimated as $F_{\text{single}}=\text{max}(P_{01} )$. We observe good agreement between the real-time simulation data and the effective Hamiltonian with RWA at low driving fields. However, at higher driving fields (e.g., $h_x=8.64$ mT), the real-time simulation shows significant deviation from the RWA, indicating its breakdown in this regime.  Additionally, the DW chirality $ C_y $ undergoes rapid oscillations during the time evolution. Notably, $ C_y $ reaches zero when $ \psi_0(t) $ evolves into the $ \psi_0 $ and $ \psi_1 $ states [see Figure~\ref{fig:S9}(b)], confirming that the time-evolving state $ \psi_0(t) $ indeed returns to either the $ \ket{0} $ or $ \ket{1} $ state.

\subsection{Two qubit gate operation}

When DW qubits pass by each other with coupling at their centers, the effective Hamiltonian is:
\begin{equation} 
    H^{\text{eff}}_{2} = \frac{\hbar \omega_{q}}{2} \left( \sigma_z \otimes I_2 + I_2 \otimes \sigma_z \right) - J^{\text{eff}}_{\text{in}} (\Delta N) \sigma_x \otimes \sigma_x,
    \label{eq:12}
\end{equation}
The interaction Hamiltonian is:
\begin{eqnarray} 
 \tilde{H}^{eff}_{2,I} &=&  - J^{eff}_{in} (\Delta N) \exp{ \left[ i \frac{  \omega_q t}{2}(  \sigma_z \otimes I_2 +  I_2 \otimes \sigma_z ) \right] } \sigma_x \otimes \sigma_x  \exp{  \left[ -i\frac{ \omega_q t}{2}(  \sigma_z \otimes I_2 +  I_2 \otimes \sigma_z ) \right] } \notag \\
&=& - J^{eff}_{in} (\Delta N)  \left(\begin{array}{cccc}0 & 0 & 0 & e^{ 2 i \omega_q t} \\0 & 0 & 1 & 0 \\0 & 1 & 0 & 0 \\ e^{- 2 i \omega_q t} & 0 & 0 & 0\end{array}\right) \notag \\
&\approx& -\frac{1}{2} J^{\text{eff}}_{\text{in}} (\Delta N)  \left( \sigma_x \otimes \sigma_x  +  \sigma_y \otimes \sigma_y \right),
    \label{eq:14}
\end{eqnarray}
where we ignore the fast oscillating term ($2 \omega_q$) in the last equation.
The unitary gate operation for two DW qubits passing by each other is:
\begin{eqnarray} 
 U_2 &=& \exp\left\{  \frac{i}{2\hbar} \int_{-\infty}^{\infty} dt \, J^{\text{eff}}_{\text{in}} (\Delta N)  \left( \sigma_x \otimes \sigma_x  +  \sigma_y \otimes \sigma_y \right)  \right\} \notag \\
 &=& \exp\left\{  \frac{i J^{\text{eff}}_{\text{in}}}{2 \hbar v_D} \int_{-\infty}^{\infty} d(\Delta N) \, \exp\left[ - (\Delta N/N_D)^2 \right]  \left( \sigma_x \otimes \sigma_x  +  \sigma_y \otimes \sigma_y \right)  \right\} \notag \\
 &=& \exp\left\{  \frac{i \sqrt{\pi} J^{\text{eff}}_{\text{in}} N_D}{ 2 \hbar v_D}  \left( \sigma_x \otimes \sigma_x  +  \sigma_y \otimes \sigma_y \right) \right\} ,
    \label{eq:15}
\end{eqnarray}
where $v_D = \Delta N / t$ is the velocity of DW qubits (units in site/s).

\subsection{Real-time simulation on two-qubit gate operation}

\begin{figure}[h]
    \centering
    \includegraphics[width=0.8\textwidth]{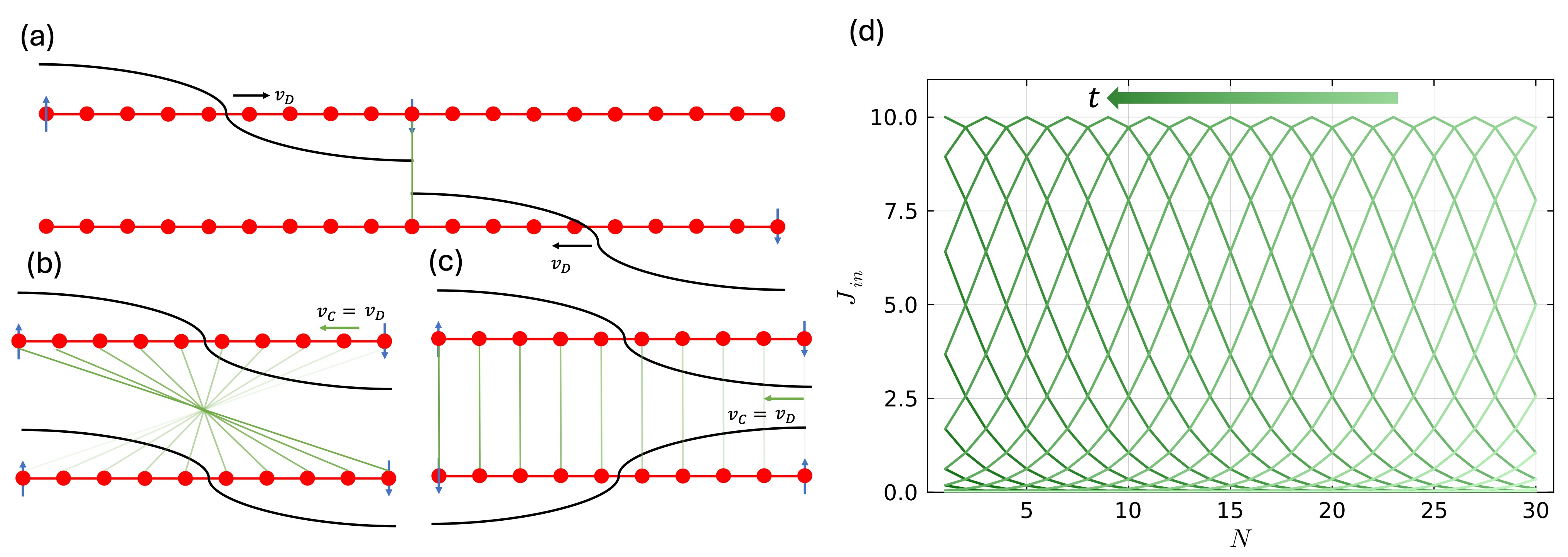}
    \caption{Schematic illustration of (a) simulating the interaction between two moving DW qubits with the interchain coupling at the center of the ladder, (b) shifting the inter-chain coupling sites with time to simulate the motion of DWs, and (c) realizing the equivalent setup with (b) by inverting spin configurations of the second DW, (d) the Gaussian profile of the time-dependent inter-chain coupling $J_i(t) $ at each time frame, with $ t_{\text{tot}} = 197.5 $ ps and $ \tau =6.58$ ps. The inset arrow denotes the direction of time flow. The width of the Gaussian profile is $\sigma = 0.2 $. 
    }
    \label{fig:S10}
\end{figure}

In Figure~\ref{fig:S10}(a), we recall the schematic plot for simulating the interaction between two moving DW qubits via a single-site coupling, as presented in the main text. Real-time simulation of such configuration requires motion of DWs on quantum spin chain. However, driving the motion of a DW in a quantum spin chain with open boundary conditions is hindered by the absence of translational invariance. Instead of directly moving the DWs, we equivalently shift the position of the single-site coupling, as illustrated in Figure~\ref{fig:S10}(b). In the initial frame, the last site of chain-1 ($N_1 = N_x$) is coupled to the first site of chain-2 ($N_2 = 1$). In a subsequent frame $i$, site $N_1 = N_x - i + 1$ of chain-1 couples with site $N_2 = i$ of chain-2. By inverting spin configurations of the second chain, this configuration becomes equivalent to a spin ladder with opposite boundary conditions, where the inter-chain coupling sites move from the right to the left in both chains at a speed of $ v_c = v_D $, as illustrated in Figure~\ref{fig:S10}(c).

When the position of single-site coupling is shifted by one lattice cite, the time-dependent Hamiltonian changes discontinuously, making real-time simulations difficult to converge. To realize adiabatic time evolution, we define the time-dependent inter-chain coupling at each site $i$
\begin{align}
J_i (t) = \tilde{J}_{in} \exp \left[ -\left( \frac{ i -  v_c (t-t_0) }{N_x \sigma} \right)^2 \right] \quad \text{for} \ i=1, 2, \cdots, N_x
\label{eq:16}
\end{align}
where $\tilde{J}_{in}$ is the strength of inter-chain coupling, $v_c$ denotes the velocity of the coupling site motion (DW velocity), $\sigma$ controls the width of inter-chain coupling, and $t_0$ is the initial condition that ensures inter-chain coupling at all site is nearly zero $J_i (0)  \approx0 $. Note that DW velocity is obtained by $v_D=v_c =N_x/(t_{tot}-2t_0)$. In Figure ~\ref{fig:S10}(d), we present the profile of $J_i (t)$ at each time step $\tau$, where the Gaussian profile of inter-chain coupling moves smoothly along the spin ladder.
The total inter-chain coupling at $t=t_{tot}/2$ is 
\begin{align}
\sum^{N_x}_{i=1} J_i (t_{tot}/2)=   \tilde{J}_{in} \sum^{N_x}_{i=1}  \exp \left[ -\left( \frac{ i - N_x/2 }{N_x \sigma} \right)^2 \right] \approx 10  \tilde{J}_{in} ,
\label{eq:17}
\end{align}
which is approximately 10 times the amplitude of the Gaussian profile. Therefore, the Gaussian profile in Eq.~\eqref{eq:16} with $\sigma = 0.2$ is approximately equivalent to a single-site coupling with a coupling strength that is 10 times smaller.

Figure~\ref{fig:S11} presents the real-time simulation of transition probabilities \( P_{ij} \) for different inter-chain coupling strengths, $ \tilde{J}_{in} $, and total simulation times $t_{tot} $. The variation in $ t_{tot} $ corresponds to different DW velocities.  In the weak coupling regime ($ \tilde{J}^\textrm{eff}_{in} \ll \Delta_{10}$), the two-qubit gate operation induced by the moving DWs is approximately an XY-gate, consistent with the effective two-qubit Hamiltonian [Figure~\ref{fig:S11}(a)] in the main text under the RWA.
For \( J_{in} = 2 \) \(\mu\)eV [Figure~\ref{fig:S11}(b)], we observe slight entanglement between the \( E_0 \) and \( E_3 \) states, suggesting that in the strong coupling regime, the two-qubit gate operation deviates from the RWA and behaves more like an XX-gate.  Examining the gate operation within the $ E_{1,2} $ subspace [Figure~\ref{fig:S11}(a1-b1, a2-b2, a3-b3)], we confirm that the required DW velocity $ v_D $ to achieve the same gate operation scales linearly with $ J_{in} $. Specifically, implementing an i-swap gate requires $ v_D \approx 11.4 $ and $ 22.8 $ sites/ns for $\tilde{J}_{in} = 1 $ and $2 $ $\mu$eV, respectively.
For both cases, the fidelity of two-qubit gates is estimated to be greater than 99\% with the leakage outside the qubit subspace below 10$^{-4}$.

\begin{figure}[h]
    \centering
    \includegraphics[width=0.8\textwidth]{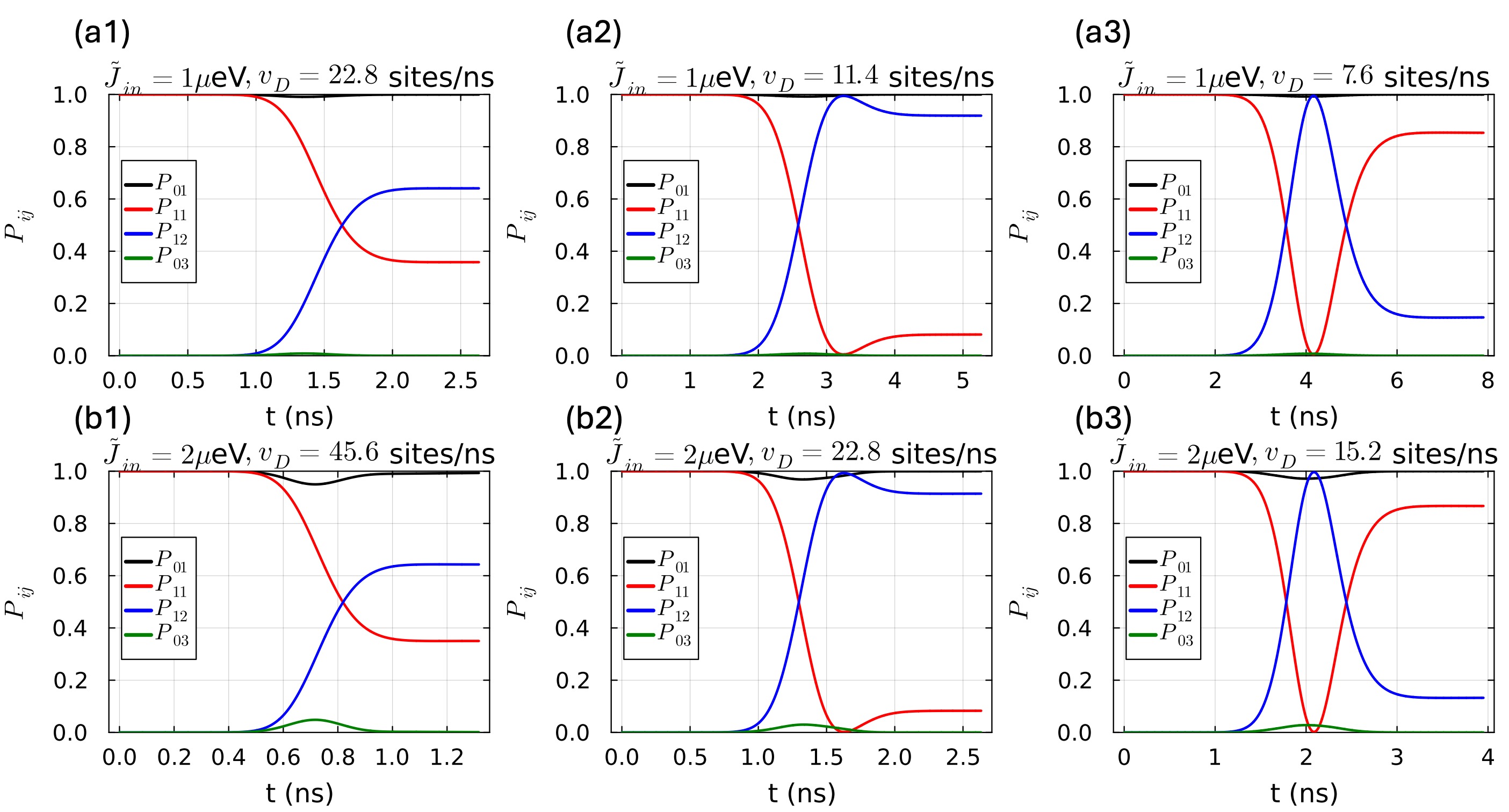}
    \caption{Comparison of real-time simulations of transition probabilities $P_{ij}$ under different inter-chain couplings $\tilde{J}_{in}$ and total simulation times $t_{tot}$.  (a1-a3) $\tilde{J}_{in} = 1$ $\mu$eV with $t_{tot} = 2.6, 5.2, 7.8$ ns. (b1-b3) $\tilde{J}_{in} = 2$ $\mu$eV with $t_{tot} = 1.3, 2.6, 3.9$ ns. The parameters for the single DW qubit are the same as in the DMRG simulations:  
$N = 30$, $J = 25.85$ meV, $K_z = 0.26$ meV, $K_y = 0.1$ meV, $h_y = 0.9$ T, and $\mu_B h_z = -100$ meV. The time step is $\tau = 1$.}
    \label{fig:S11}
\end{figure}

\section{Quantum computing platform of domain wall qubits}

\subsection{Domain wall motion in magnetic insulators}

The decoherence of DW qubits results from interactions with other particles, including electrons, phonons, and magnons, which is characterized by a phenomenological Gilbert damping constant, denoted as $\alpha$.
Since the electron scattering often results in a large damping constant in metallic systems~\cite{kamberskySpinorbitalGilbertDamping2007}, magnetic insulators are desirable for quantum applications.
DWs in insulators can be driven by spin-orbit torque~\cite{shaoRoadmapSpinOrbit2021}, which arises from the spin Hall effect at the interface between ferromagnets and heavy metals~\cite{hirschSpinHallEffect1999}.
Importantly, the chirality of N\'{e}el-type DW plays a crucial role in the spin-orbit torque-driven motion.
While DWs with opposite chiralities move in opposite directions, those with the same chirality undergo unidirectional motion~\cite{haazenDomainWallDepinning2013, emoriCurrentdrivenDynamicsChiral2013}.
In contrast, the spin-orbit torque cannot drive Bloch-type DWs due to the rotational symmetry about the $y$ axis.

\begin{figure}[h]
    \centering
    \includegraphics[width=0.8\textwidth]{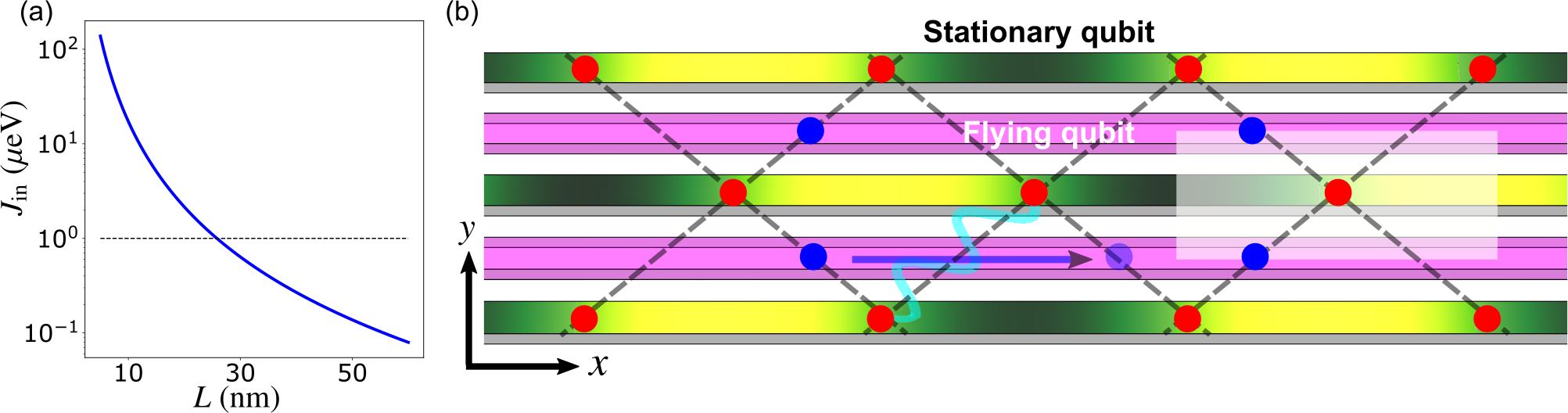}
    \caption{(a) Effective interchain coupling strength between two DWs is plotted as a function of their distance at $N_s=200$. The dashed line indicates $J_\textrm{in}=1~\mu$eV, which is obtained at $L\approx 26~$nm.
    (b) Schematic illustration of two-dimensional architectures of DW qubits. Stationary qubits~(red circles) are entangled via flying qubits (blue circles). Dashed lines indicate the connectivity of qubits, which form a square lattice with nearest-neighbor interactions. A filled square denotes the physical size of DW qubits, estimated as $75\times 150$~nm$^2$ when the width and thickness of nanowires are 10~nm and 1~nm, respectively. 
    }
    \label{fig:S12}
\end{figure}

\subsection{Magnetic dipole-dipole interaction as interchain coupling}
As discussed in the main text, the interchain exchange coupling leads to the two-qubit interaction between DW qubits.
However, the connectivity of qubits is limited for short-range direct exchange coupling.
Here, we consider the long-range magnetic dipole-dipole interaction as the effective interchain coupling. We note that the Ruderman–Kittel–Kasuya–Yosida (RKKY)-type interaction via conduction electrons could also result in the interchain coupling~\cite{rudermanIndirectExchangeCoupling1954, kasuyaTheoryMetallicFerro1956, yosidaMagneticPropertiesCuMn1957}.
The magnetic dipole-dipole interaction between a pair of spins is given by
\begin{equation}
	H_\textrm{dipole}(\boldsymbol{r},\boldsymbol{r}')
	=
	-\frac{\mu_0 \gamma_e^2\hbar^2}{4\pi|\boldsymbol{r}'-\boldsymbol{r}|^3}
	\left[\frac{3\{\boldsymbol{S}_{\boldsymbol{r}}\cdot(\boldsymbol{r}'-\boldsymbol{r})\}\{\boldsymbol{S}_{\boldsymbol{r}'}\cdot(\boldsymbol{r}'-\boldsymbol{r})\}}{|\boldsymbol{r}'-\boldsymbol{r}|^2}-\boldsymbol{S}_{\boldsymbol{r}}\cdot\boldsymbol{S}_{\boldsymbol{r}'}\right],
\label{eq_dipole}
\end{equation}
where $\gamma_e$ is the gyromagnetic ratio, and $\boldsymbol{S}_{\boldsymbol{r}}$ and $\boldsymbol{S}_{\boldsymbol{r}'}$ denote dimensionless spin vectors of two nanowires at $\boldsymbol{r}$ and $\boldsymbol{r}'$, respectively.  
The effective magnetic field at centers of DWs is given  as
\begin{align}
	\boldsymbol{H}_\textrm{eff}(\boldsymbol{r}_\textrm{c})
	&=
	\sum_{\boldsymbol{r}'}
	\frac{\partial H_\textrm{dipole}(\boldsymbol{r}_c,\boldsymbol{r}')}{\partial \boldsymbol{S}_{\boldsymbol{r}_c}}
	=
	-\sum_{\boldsymbol{r}'}\frac{\mu_0 \gamma_e^2\hbar^2}{4\pi|\boldsymbol{r}'-\boldsymbol{r}_c|^3}
	\left[\frac{3\boldsymbol{S}_{\boldsymbol{r}'}\cdot(\boldsymbol{r}'-\boldsymbol{r}_c)}{|\boldsymbol{r}'-\boldsymbol{r}_c|^2}(\boldsymbol{r}'-\boldsymbol{r}_c)-\boldsymbol{S}_{\boldsymbol{r}'}\right]
	\\
	&\approx
	-
	\frac{\mu_0 \gamma_e^2\hbar^2N_s}{4\pi L^3}
	\left[3(\boldsymbol{S}_{\boldsymbol{r}'_c}\cdot\hat{\boldsymbol{y}})\hat{\boldsymbol{y}}-\boldsymbol{S}_{\boldsymbol{r}'_c}\right]
	\approx
	-\frac{\mu_0 \gamma_e^2\hbar^2 N_s}{2\pi L^3}\boldsymbol{S}_{\boldsymbol{r}'_\textrm{c}},
\end{align}
where $N_s$ is the number of spins within DWs, $\boldsymbol{r}_c$ and $\boldsymbol{r}'_c$ denote the position of DW centers in each nanowire, respectively.
Since the $z$ component of spins is antisymmetric about the center of DWs, the contribution of spins away from DW centers vanishes.
Consequently, the distance between two nanowires is approximated as $\boldsymbol{r}'-\boldsymbol{r}_c\approx L\hat{\boldsymbol{y}}$.
Furthermore, we simplify the expression by assuming $(\boldsymbol{S}_{\boldsymbol{r}'}\cdot\hat{\boldsymbol{y}})\hat{\boldsymbol{y}}\approx\boldsymbol{S}_{\boldsymbol{r}'} $, which is exact at the center of DWs for $h_x=0$.
Finally, we obtain the effective interchain coupling due to the magnetic dipole-dipole interaction as 
\begin{equation}
H_\textrm{inter}\approx -J_\textrm{in} \boldsymbol{S}_{\boldsymbol{r}}\cdot \boldsymbol{S}_{\boldsymbol{r}'},
\end{equation}
with $J_\textrm{in}=\mu_0 \gamma_e^2\hbar^2 N_s^2/(2\pi L^3)$.
Assuming the width and thickness of nanowires as 10~nm and 1~nm, respectively, the total number of spins is estimated as $N_s=200$. 
Figure~\ref{fig:S12}(a) shows the effective interchain coupling strength $J_\textrm{in}$ as a function of the displacement $L$.
The interchain coupling strength is obtained as $1~\mu$eV at $L\approx 26~$nm.

\subsection{Two-dimensional quantum platform}

\paragraph{Basic architecture}
We propose two-dimensional arrays of DW qubits as a scalable quantum platform, where DW qubits act as both stationary qubits and flying qubits to improve their connectivity.
Figure~\ref{fig:S12}(b) illustrates this concept, where stationary qubits are arranged in a square lattice and connected by flying qubits.
In this setup, a $(M+1)\times (M+1)$ square lattice requires $4M+1$ nanowires.
Since the spin-orbit torque-driven motion depends on the chirality of DWs~\cite{haazenDomainWallDepinning2013, emoriCurrentdrivenDynamicsChiral2013}, we can control the motion of individual DW qubits by locally manipulating their chiralities.
Hence, it is possible to perform two-qubit gates between selected pairs of nearest neighbors with a single current pulse.
Furthermore, entangling remote qubits can be achieved by precisely controlling the position and velocity of DWs, which is proportional to the pulse duration and current density~\cite{taniguchiPreciseControlMagnetic2015}.
The position of DWs can be also fixed by modulating the thickness of heavy metal layers~\cite{leePositionErrorfreeControl2023}.

\paragraph{Initialization}
DW qubits can be created in magnetic nanowires by various methods such as magnetic fields or current injections~\cite{parkin2008magnetic}.
The qubit can be initialized to the $\ket{0}$ state by applying the magnetic field along the $x$ axis, thereby fixing the chirality of DWs. 
We can also control the chirality by tuning the interfacial Dzyaloshinskii–Moriya interaction via gate voltages~\cite{fillionGatecontrolledSkyrmionDomain2022}. 

\paragraph{Measurement}
DW qubits can be measured in the chirality basis using nanoscale imaging techniques.
The spatial resolution of the NV magnetometry ranges from 10~nm to 50~nm~\cite{thielQuantitativeNanoscaleVortex2016, songDirectVisualizationMagnetic2021}, holding promise for direct imaging of DWs.
Another idea is to employ measurement techniques of spin qubits, where the polarization of electrons near the center of DW qubits is measured to determine the chirality of DWs~\cite{zou_quantum_2023}.
Furthermore, the chirality of DWs affects the direction of spin-orbit torque-driven motion.
Thus, the chirality could also be measured by monitoring the DW motion with arrays of anomalous Hall detectors~\cite{jeonMulticoreMemristorElectrically2024}. Although the readout time has not been estimated in the present study, we note that it can be tailored through the use of quantum error correction codes, at the cost of reduced fidelity~\cite{PhysRevA.109.032433}.

\paragraph{Single-qubit gate}
The Pauli-X gate can be implemented by local in-plane magnetic fields using on-chip striplines~\cite{zhangHighlyEfficientDomain2016}.
Furthermore, all-electrically driven Rabi rotations can be realized by gate voltages in metal-oxide/ferromagnet/metal trilayer systems~\cite{fillionGatecontrolledSkyrmionDomain2022}.
For the Pauli-Z gate, we can apply continuous microwave magnetic fields along the $y$ axis while locally tuning the resonance frequencies of qubits. A similar method was successfully demonstrated in spin qubits, achieving the fidelity above 99\%~\cite{lauchtElectricallyControllingSinglespin2015}.

\paragraph{Two-qubit gate}
The CNOT gate can be performed by shuttling two DW qubits towards each other, as discussed in the main text.
The fidelity of two-qubit gates relies on the precise control of DW velocity, which is linearly proportional to the current density above the critical current density and remains almost constant regardless of current pulse durations~\cite{taniguchiPreciseControlMagnetic2015}.
Therefore, it is crucial to minimize noise in current pulses for achieving high fidelity in CNOT gates.

\paragraph{Scalablility}
The width of magnetic nanowires typically ranges from 100~nm to 500~nm in the study of racetrack memory~\cite{parkin2008magnetic}.
However, sub-100-nm-wide wires can be fabricated by high-resolution patterning techniques~\cite{duttaMicromagneticModelingDomain2015}.
Recent studies also showed that the DW motion can be tracked electrically with a spatial resolution below 40~nm~\cite{jeonMulticoreMemristorElectrically2024}, which establishes a lower limit for the spacing between DW qubits.
We estimate the optimal horizontal spacing between stationary qubits and flying qubits to be $\sqrt{2}L$ as shown in Fig.~\ref{fig:S12}(b), where the $y$ component of the effective magnetic field from the dipolar interaction vanishes between them. 
In this setup, the physical size of DW qubits is $75\times 150$~nm$^2$, comparable to the size of spin qubits.
With $10~\mu$m long nanowires, a square lattice of $66\times 66\approx 4300$ qubits could be realized within $10\times 10~\mu$m$^2$.

\paragraph{Error correction}
We suggest the implementation of surface codes as a possible quantum error correction scheme~\cite{fowlerSurfaceCodesPractical2012}.
Crucially, a simple square lattice of qubits with nearest-neighbor interactions is sufficient for surface codes.
Another advantage of surface codes is their high fault-tolerance threshold, which reaches 1\% under certain conditions~\cite{raussendorfFaultTolerantQuantumComputation2007, fowlerPracticalClassicalProcessing2012}.
Although many physical qubits are required for logical qubits of surface codes, the two-dimensional architecture of DW qubits provides an attractive platform with exceptional scalability.

 \begin{table}[t]
\begin{tabular}{lcccc}\hline\hline
Type   & $T_1,T_2$   & Physical size  &  Gate rate (2Q) & Quality factor \\
  \hline\hline
Ion qubits~\cite{10.1063/1.5088164} 
&0.1 s\,$\sim$\,100~s &$100\times 100~\mu$m$^2$ & $\sim$ 10 kHz & $ 10^6$
\\ 
Superconducting qubits~\cite{kjaergaardSuperconductingQubitsCurrent2020} 
&10~ns\,$\sim$\,1~ms &$100\times 100~\mu$m$^2$ & $\sim$ 10 MHz & $ 10^4$
\\ 
Spin qubits~\cite{stanoReviewPerformanceMetrics2022}
& 1~ns\,$\sim$\,57~s & $100\times 100$~nm$^2$ &  $\sim$ 0.1 GHz & 500
\\
DW qubits (this work)~\cite{zou_quantum_2023}
& 0.2~$\mu$s \,$\sim$\,30~$\mu$s  & $75\times150$~nm$^2$ & 1 GHz & 2$\times 10^3$\,$\sim$\, 3$\times 10^5$
\end{tabular}
\caption{\label{tab:1} Comparison of different types of qubits. In the column of coherence time $T_1$ and $T_2$, we cite values from the early work to the latest results to illustrate the remarkable improvement over recent years.
The coherence time of DW qubits is estimated with the lowest reported value $\alpha=1\times 10^{-5}$ and the linear fitted value $\alpha=6\times10^{-8}$ at 50~mK~\cite{maier-flaigTemperaturedependentMagneticDamping2017}.  Gate rate (2Q) denotes the inverse of two-qubit gate operation time.
 Quality factor denotes the number of Rabi oscillations within the coherence time. 
}
\end{table}

\paragraph{Comparison with other platforms }
Table~\ref{tab:1} summarizes the quantitative comparison of DW qubits with other quantum computing platforms. The physical size of DW qubits is comparable to that of spin qubits, which reemphasizes the potential compatibility and integrability of these two quantum platforms. The gate rate of DW qubits, reaching 1 GHz, surpasses that of all other quantum computing platforms, indicating their high potential for realizing high-clock-speed quantum processors.  Although the current limitation of DW qubits lies in their short coherence time, it is worth highlighting the significant increase in the coherence time of superconducting qubits and spin qubits over recent decades.
Similarly, the coherence time of DW qubits could be improved substantially by material engineering.
The coherence time of DW qubits is inversely proportional to the Gilbert damping constant~\cite{zou_quantum_2023}, whose lowest value was reported at room temperature~\cite{klinglerGilbertDampingMagnetostatic2017}.
Thus, the Gilbert damping constant could be several orders of magnitude smaller at the cryogenic condition.
According to the magnon-phonon scattering theory~\cite{kasuyaRelaxationMechanismsFerromagnetic1961}, the Gilbert damping is predicted to decrease linearly with temperatures.
The experimental study in YIG spheres confirms this linear dependence above 100~K~\cite{maier-flaigTemperaturedependentMagneticDamping2017}, where we extrapolate $\alpha\approx 6\times10^{-8}$ at 50~mK.
This work indicates that the ultrasmall damping could be realized in a clean magnetic sample.
In Table~\ref{tab:1}, we cite the coherence time of DW qubits from Ref.~\onlinecite{zou_quantum_2023} with the lowest reported value $\alpha=1\times 10^{-5}$ and the linear fitted value $\alpha=6\times10^{-8}$ to highlight the potential of DW qubits in comparison to existing platforms.

\end{widetext}

\end{document}